\documentclass[aps,prd,twocolumn,preprintnumbers,nofootinbib]{revtex4-1}

\usepackage{graphicx}
\usepackage[FIGTOPCAP]{subfigure}
\usepackage{dcolumn}
\usepackage{bm}
\usepackage{amsmath}
\usepackage{epstopdf}
\usepackage{amsmath,amssymb}
\usepackage{bigints}
\usepackage{mathtools}
\usepackage{array}
\usepackage{bbm}
\usepackage{dsfont}
\usepackage{todonotes}
\usepackage[normalem]{ulem} 
\usepackage[breaklinks=true,colorlinks,citecolor=blue,linkcolor=blue,urlcolor=blue]{hyperref}

\newcommand{\beq}{\begin{eqnarray}}
\newcommand{\eeq}{\end{eqnarray}}

\newcommand{\real}{{\sf I}\kern-.12em{\sf R}}
\newcommand{\comp}{{\sf I}\kern-.50em{\sf C}}
\newcommand{\unity}{{\sf I}\kern-.54em{\sf 1}}

\newcommand{\mcTau}{\mathcal{T}}
\newcommand{\YMconfs}{\mathcal{C}_{G}\big( \mcTau \big)}

\begin{document}

\title{Compact Gauge Fields on Causal Dynamical Triangulations: 
a 2D case study}

\author{Alessandro Candido}\email{alessandro.candido@mi.infn.it}
\affiliation{Dipartimento di Fisica dell’Università degli Studi di Milano and INFN 
    - Sezione di Milano,\\ Via Giovanni Celoria, 16 - 20133 Milan, Italy}
\author{Giuseppe Clemente}\email{giuseppe.clemente@pi.infn.it}
\author{Massimo D'Elia}\email{massimo.delia@unipi.it}
\affiliation{Dipartimento di Fisica dell'Universit\`a di Pisa and INFN
	- Sezione di Pisa,\\ Largo Pontecorvo 3, I-56127 Pisa, Italy.}
\author{Federico Rottoli}\email{frottoli@sissa.it}
\affiliation{SISSA -- International School for Advanced Studies, \\
    via Bonomea 265,  34136 Trieste, Italy.}

\date{\today}

\begin{abstract}
We discuss the discretization of Yang-Mills theories on 
Dynamical Triangulations in the compact formulation,
with gauge fields living on 
the links of the dual graph associated with the triangulation,
and the numerical investigation
of the minimally coupled system by Monte Carlo simulations. 
We provide, in particular, an explicit construction 
and implementation
of the Markov chain moves for 2D Causal Dynamical Triangulations 
coupled to either $U(1)$
or $SU(2)$ gauge fields; the results of exploratory numerical simulations
on a toroidal geometry are also presented for both cases. 
We study the critical behavior 
of gravity related observables, determining the associated critical indices,
which turn out to be independent of the bare gauge coupling:
we obtain in particular $\nu = 0.496(7)$ for the critical index regulating 
the divergence of the correlation length of the volume profiles.
Gauge observables are also investigated, including holonomies (torelons)
and, for the $U(1)$ gauge theory, the winding number and 
the topological susceptibility.
An interesting result is that the critical slowing down of the 
topological charge, which affects various lattice field theories
in the continuum limit, seems to be strongly suppressed (i.e.~by orders
of magnitude) by the presence
of a locally variable geometry: that may suggest possible
ways for improvement also in other contexts.
\end{abstract}

\maketitle

\section{Introduction}

Dynamical Triangulations represent one of the open possibilities for a  
solution to the quest for a self-consistent Quantum Theory of Gravity.
The approach is based on the so-called asymptotic safety 
program~\cite{ass_weinberg}, which has the aim of providing 
a renormalized theory by finding a non-perturbative UV fixed point 
in the space of parameters. A possibile implementation of such 
program is numerical and based on the exploration, by 
path-integral Monte Carlo methods,
of the phase diagram of the discretized theory, looking for 
critical points in the parameter space, which are candidate 
UV fixed points where continuum physics could be investigated.
A standard discretization is the one based on the Regge formalism~\cite{regge}, 
where space-time configurations are represented by triangulations,
i.e., collections of flat simplexes glued together according to 
different geometries. Promising results have been obtained, in particular,
by exploring Causal Dynamical Triangulations (CDT)~\cite{cdt_pioneer,cdt_review12,cdt_review19,DynTriLor,cdt_secondord,cdt_secondfirst,new_phase_chars,cdt_newhightrans,cdt_toroidal,cdt_quantum_Ricci_curv,cdt_phasestruct_toroidal,LBseminal,LBrunning,cdt_higherord_toroidal,cdt_quantum_Ricci_curv_round,LBFEMseminal}, 
where one enforces an additional condition of global hyperbolicity 
by means of a space-time foliation~\cite{causconds}.

The properties of Dynamical Triangulations in the presence of additional
quantum fields, e.g., matter or gauge fields, has been considered various 
times in the past. This is interesting for at least two reasons.
First, if Dynamical Triangulations eventually reveal successful
in describing a continuum, renormalized theory of gravity, then 
lattice numerical
simulations of gravity coupled to matter and gauge fields 
will represent the appropriate setting for the investigation 
of the quantum cosmological regime. 
Second, considering additional quantum fields could be essential
even in the very process of searching for an UV fixed point, since it 
enlarges the coupling space and the number of candidate continuum theories.

In this paper we focus on the addition of gauge fields, either Abelian or 
non-Abelian. This has been considered at an analytic level in
the past literature~\cite{cdtgauge_anal}, or numerically 
by considering a non-compact formulation
of gauge fields~\cite{cdtgauge_noncompact}.
Our study is devoted to a full numerical
implementation based on compact gauge fields:
to that purpose, after a discussion about 
the formulation on generic triangulations,
we focus on the development of the 
Markov chain moves and on actual numerical simulations
for a simplified
two-dimensional model, in particular 2D CDT, adding to it
either $U(1)$ or $SU(2)$ gauge fields by minimal coupling. 
One advantage of the model is
that it is known to be renormalizable, so that one has the possibility
to analyze how the critical behavior 
of the system is modified by the additional coupling.

One important aspect concerns the way the gauge symmetry is implemented 
in the context of Dynamical Triangulations. We adopt the point of view 
according to which the local gauge is fixed consistently over each
simplex: in this way, elementary parallel transports, 
which substitute the continuum gauge connection in lattice gauge theories, 
are naturally associated with the links of the dual graph,
i.e., links connecting adiacent symplexes, which are dual to 
the hypersurfaces separating them. We consider this as the most
natural framework, since the choice of the local reference frame 
is associated, both for internal and Lorentz symmetries, 
with the same local object, i.e., the symplex; moreover, this is 
also consistent with the correct counting for the 
density of gauge degrees of freedom~\cite{cdtgauge_noncompact}. 

We notice that, in the particular case of 2D triangulations, there 
is also an equivalent choice, 
frequently adopted in past studies~\cite{cdtgauge_anal}, 
which associates gauge links with 
the actual links of the triangulation, i.e., the edges of the triangles,
and gauge transformation with the triangle vertices. However, this possibility
is left open only in two dimensions, due to the fact that, in this case,
the hypersurfaces delimiting a simplex are one-dimensional objects
and the number of simplexes equals two times the number of vertices. 
Since we consider the present study as a pilot exploration in view
of a generalization to a larger number of dimensions, we keep the point of 
view discussed above.

The paper is organized as follows.
In Section~\ref{sec:mincoup} we will describe the action
of Yang-Mills theory minimally coupled to gravity and how
we discretize it. The set of Markov chain moves adopted in 
our Monte Carlo simulations is described schematically in 
Section~\ref{sec:algo}, while a detailed discussion about 
the detailed balance conditions is postponed to the Appendix.
In Section~\ref{sec:numres} we will present the results of 
our numerical simulations and finally, in Section~\ref{sec:conclusions}
we report our conclusions.

\section{Minimal coupling of Yang-Mills theories to CDT}\label{sec:mincoup}

In the following, we will denote by $\YMconfs$ the space of all 
possible gauge field configurations on the triangulation $\mcTau$, and with gauge group $G$.
The action for the composite gravity-gauge system can be formally
decomposed, in the minimal coupling paradigm, 
as $S[\Phi,\mcTau] = S_{CDT}[\mcTau] + S_{\text{YM}}[\Phi;\mcTau]$, 
with $\Phi \in \YMconfs$ and where $S_{\text{YM}}$ represents the minimally 
coupled action of the field $\Phi$ in the backgroud $\mcTau$.
We now have to derive the specific form which $S_{\text{YM}}$ has to take in order to 
represent correctly the continuum Yang-Mills theory; as a guide to this task, we will
use the analogy with the flat background case.

\subsection{Yang-Mills theories on a flat lattice}
In the continuum, Yang-Mills field configurations can be represented by $1$-forms $A_{\mu}$ 
with values in the Lie algebra of the gauge group, $\text{Lie}(G)$,
while the Yang-Mills action in the Euclidean space 
takes the form: 
\begin{equation}\label{eq:un-YMact_flat}
    S_{\text{YM}} = \frac{1}{4} \bigintssss\limits_{\mathcal{M}} \! d^d x \; F^{a}_{\mu \nu} F^{a \mu \nu}
,
\end{equation}
with $F^{a}_{\mu \nu}=\partial_\mu A^{a}_{\nu} - \partial_\nu A^{a}_{\mu} + g f^{abc} A^{b}_{\mu} A^{c}_{\nu}$,
and $A_{\mu}=A^{a}_{\mu} T^{a}$, with $T^{a}$ the generators of the Lie algebra ($a\in\{1,\dots,N^2-1\}$ for $SU(N)$, with the standard normalization
  ${\rm Tr}(T_a T_b) = \delta_{ab} /2$).

This theory is usually discretized on a flat cubic lattice in terms of gauge link
variables $U_\mu(n)$, i.e., the elementary parallel transporters, with values
in the gauge group, linking adiacent sites of the lattice.
The continuum action can be discretized in various different ways,
the simplest one is the so-called {\em plaquette} (or Wilson) action:
\begin{equation}\label{eq:un-YMact}
S_{\text{YM}} \equiv - \frac{2N}{g^2} \sum\limits_{\Box} \Big[ \frac{1}{N} Re Tr \Pi_\Box - 1 \Big] \, ,
\end{equation}
where $\Pi_\Box$ is the oriented product of gauge link variables around 
an elementary plaquette $\Box$ of the lattice, and the sum extends over 
all possible plaquettes; for 
the $U(1)$ gauge theory, 
$2N/g^2$ is substituted by $1/g^2$.

\subsection{Yang-Mills action coupled to CDT triangulation}
The continuum version of the action of a Yang-Mills theory minimally 
coupled to gravity (described by the Einstein-Hilbert action) 
reads as follows:
\begin{equation}\label{eq:un-YMEHact_flat}
    S_{\text{YM + EH}} = 
\bigintssss\limits_{\mathcal{M}} \! d^d x \sqrt{-g} \; 
\left[ \frac{1}{4} F^{a}_{\mu \nu} F^{a \mu \nu} + (R - 2 \Lambda) \right]
,
\end{equation}
where $R$ is the scalar curvature, $\Lambda$ is the cosmological constant
and the $g$ in $\sqrt{-g}$ stands in this case
for the determinant of the metric tensor. 
In two dimensions, which is the specific case considered in this study, 
the Gauss-Bonnet theorem makes the curvature term only dependent 
on the global topology of the manifold: since the latter is usually
kept fixed in CDT simulations (e.g., toroidal), such term 
can be ignored, leaving the cosmological constant as the only 
relevant coupling in the gravity sector.

Indeed, the pure-gravity action contribution for CDT in two dimensions,
after Wick rotation to the Euclidean space, has the simple expression~\cite{cdt_review12}:
\begin{equation}\label{eq:mincoup-CDT2Daction}
    S_{CDT,2D}[\mcTau]= \lambda N_2[\mcTau],
\end{equation}
where $N_2[\mcTau]$ is the number of triangles in the triangulation $\mcTau$ 
(i.e., the total volume), 
and $\lambda$ is the only free parameter, which is related to
the cosmological constant. When considering the path-integral 
formulation of pure CDT, i.e., the integral over all possible triangulations weighted 
by $\exp( - S_{CDT,2D}[\mcTau])$, one observes a critical behavior 
as $\lambda \to \lambda_c = \log 2$ from above, where both the average 
global volume and the correlation length for foliation volumes 
diverge.

\begin{figure}
\centering
\includegraphics[width=0.5\textwidth]{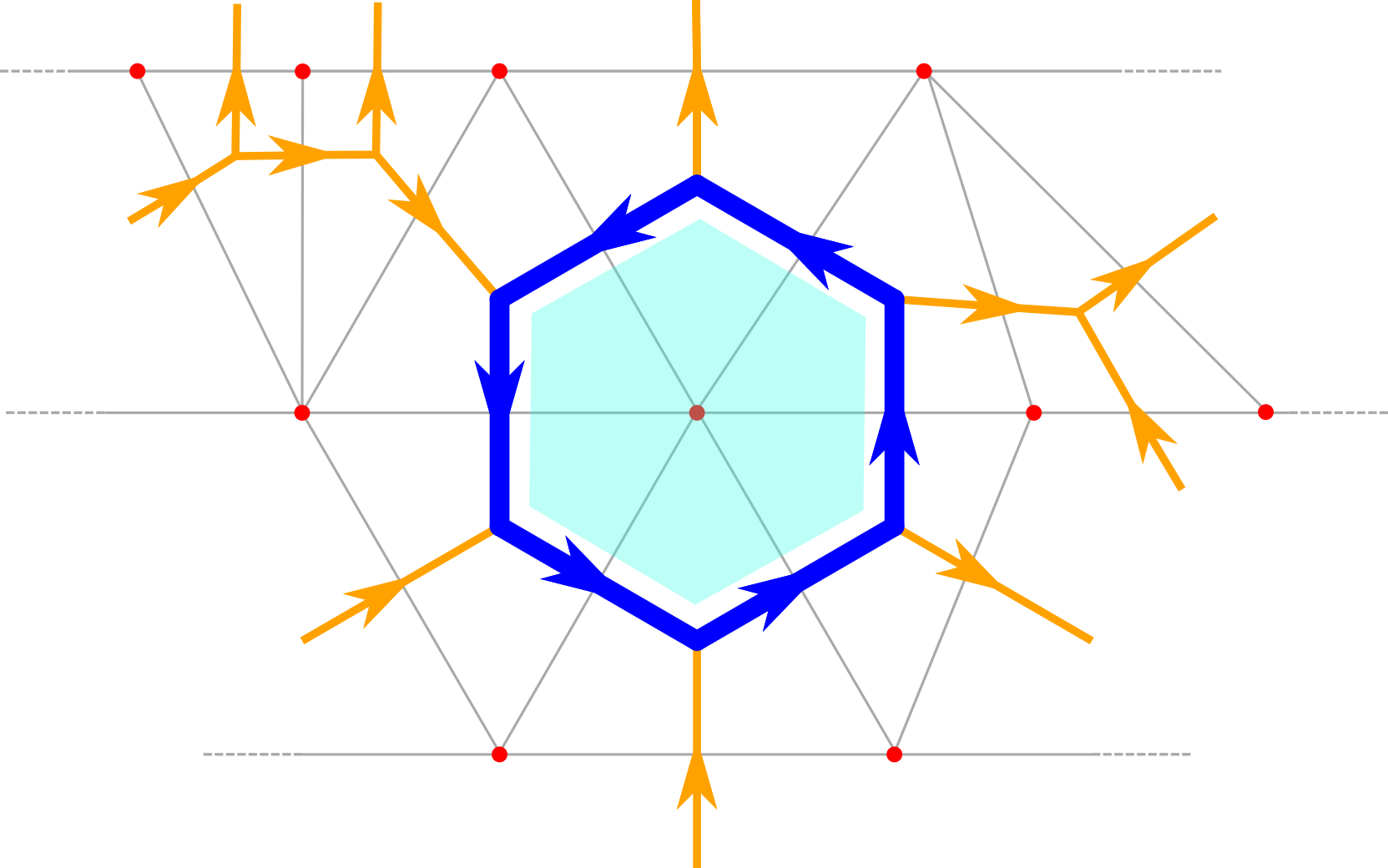}
\caption{Sketch of a gauge plaquette on a typical 2D CDT triangulation. 
For simplicity, we have considered a plaquette around a vertex with zero 
cruvature, i.e., $n_b = 6$. The direction of gauge links not belonging 
to the plaquette has been drawn according to the convention adopted in
our code, i.e., from left to right and from bottom to up.
Even if drawn differently, due to the representation on a plane, 
all triangles in this and the following figures 
should be considered equilateral.}
\label{fig:plaquette_draw}
\end{figure}

Let us now consider the discretization of gauge fields. As already 
explained in the introductory discussion, we assume that a different
choice of the local gauge is associated with each single simplex. 
Therefore, the information about the gauge connection is carried
over by the elementary parallel transporters connecting each couple 
of adiacent simplexes: the gauge links connect ideally the centers 
of the simplexes, so they all have equal elementary length\footnote{This
is true in the present case, but
can be modified if an anisotropy is introduced in the definition of 
the simplexes, as usual in 4D CDT.} and are naturally associated
with the links of the dual graph. 

For a 2D causal triangulation, like that
sketched in Fig.~\ref{fig:plaquette_draw}, one can easily describe 
the gauge configuration as $U_\mu(s)$, where $s$ stands for a particular
simplex (the analogous of a lattice site in standard lattice gauge theories)
and $\mu$ is either temporal or spatial, with positive orientations 
taken upward in time and rightward in space; each simplex is associated
with two spatial links (one ingoing and one outgoing) and 
with just one temporal link (either ingoing or outgoing).
Of course this structure is specific to causal triangulations, while 
the general formulation described in the previous paragraph 
applies to general (including non causal) triangulations.

In this formulation, the elementary gauge invariant objects are associated
with traces of elementary closed loops over the dual graph: an example is shown
in Fig.~\ref{fig:plaquette_draw} for the 2D case. In analogy with 
standard lattice gauge theories, the name \emph{plaquette} will be used
for these elementary closed loops.
 A plaquette encloses an elementary 
two-dimensional surface of the dual graph, which is dual to a 
$(d-2)$ simplex of the original triangulation: this is  
usually known as a \emph{bone} in the Regge terminology~\cite{regge},
and is the geometrical object where curvature resides in the
Regge formulation. In the dual graph representation, 
it becomes the object where the \emph{gauge curvature} resides as well.

A plaquette $\Pi_b$ centered around bone $b$ will
then correspond to the ordered product 
of $n_b$ dual link variables around $b$,
where $n_b$ is the coordination number of the bone $b$.
In the particular example shown in 
Fig.~\ref{fig:plaquette_draw}, $n_b = 6$. We remind 
the reader that for a two-dimensional triangulation 
$n_b = 6$ corresponds to a locally flat space-time, 
while $n_b < 6$  and $n_b > 6$ correspond, respectively, 
to positive and negative local curvature.

The area enclosed by an elementary plaquette is 
$A n_b$, where $A$ is an elementary unit of area (one third of the 
simplex area in two dimensions). Therefore, it is 
easy to write down the correspondence between plaquettes 
and continuum field strengths in the na\"ive continuum limit:
\begin{equation}
\Pi_b \simeq \exp\left(i g n_b A F_{\mu\nu}\right)
\end{equation}
where $\mu$ and $\nu$ define the directions orthogonal to the bone.
When one takes the trace of the plaquette, in order to obtain 
a gauge invariant contribution, the second order expansion
of the previous expression gives origin to the standard 
$F_{\mu\nu}^2$ contribution in the action; however this is 
accompained by a $n_b^2$ factor which, contrary to standard lattice 
gauge theories where the lattice geometry is fixed, is a dynamical 
variable that must be properly taken into account,
 as we discuss in the following.

In the continuum action, $F_{\mu\nu}^2$ stands in front of the 
$d^d x \sqrt{-g}$ factor, which is a measure of the 
physical volume. The volume can be counted over the triangulation 
in many different ways: one obvious way is to sum over all symplexes
counting an elementary volume for each of them;
an alternative way, which is useful for our purposes, is to sum 
over all bones $b$, taking into account that the volume
around each bone is proportional to the number
of simplexes sharing the same bone, i.e., to the coordination number $n_b$.
From this reasoning, it is clear that an action which is given
as a sum over bones, as the plaquette action in our case, must include
a factor $n_b$ for each bone. Since $F_{\mu\nu}^2$ 
comes out of the plaquette with a $n_b^2$ factor attached, it is necessary
to divide by $n_b$ the contribution from each plaquette.

To summarize, the form of the plaquette action over 
a triangulation $\mcTau$ can be written, apart from 
a global factor, as follows
\begin{equation}\label{eq:un-YMact_cdt}
    S_{\text{YM}} \equiv - \beta\!\!\!\! \sum\limits_{b \in \mcTau^{(d-2)}}\!\! \frac{\widetilde{\Pi}_b}{n_b},
\end{equation}
where we use the shorthand 
\mbox{$\widetilde{\Pi}_b\equiv \lbrack \frac{1}{N} Re Tr \Pi_b - 1 \rbrack$}, 
the set of all bones of the triangulation $\mcTau$ is denoted by $\mcTau^{(d-2)}$
(which are $(d-2)$-simplexes) and $\beta$ is proportional to $1/g^2$.
This expression is valid for any dimension $d$.

\section{Algorithm for 2D CDT}\label{sec:algo}

The Monte Carlo computation of path-integral averages in the 
discretized theory requires to sample the possible 
triangulation+gauge field configurations according to 
the path-integral weight $\exp(-S_{CDT,2D} - S_{\text{YM}})$, 
with action contributions given by Eqs.~\eqref{eq:mincoup-CDT2Daction} 
and~\eqref{eq:un-YMact_cdt}.
This is achieved, as usual, by a set of Markov chain moves
which must be properly selected in order to ensure
detailed balance, ergodicity and aperiodicity.
Since the action is local, a convenient set of 
moves is given by elementary steps that 
change locally the triangulation, the gauge configuration or both.

In pure gravity CDT simulations, one needs ergodic moves that 
modify the triangulation without spoiling the foliation,
that is, the move must modify the geometry while maintaining the
causal structure. A commonly used set are the Alexander moves~\cite{alexander,cdt_2dmoves}, 
which in two dimensions are denoted by $(2,2)$, $(2,4)$ and
$(4,2)$, as the number of triangles involved before and after the move: 
they are Metropolis--Hastings steps~\cite{metro,hastings}, which respectively 
reorder, create or destroy a subset of simplexes and that
we will review later in this section.
When we consider gauge field coupled to gravity,
the space of gauge configurations $\YMconfs$ depends also on the 
triangulation $\mcTau$. The action of the Alexander moves,
thus, have also an effect on the gauge field, including the 
creation or destruction of gauge links, that must be properly
taken into account.

Moreover, we require a set of moves ensuring
that the Markov chain can ergodically explore the configuration
space of gauge fields $\YMconfs$ at fixed geometry, that is, by 
changing the gauge configuration while 
maintaining the triangulation invariant.

In the following, we provide a brief description of the
local moves implemented in our Markov chain Monte Carlo algorithm.
A more detailed discussion about detailed balance for such moves
is reported in the Appendix.

\subsubsection{Pure gauge move}\label{subsubsec:un-db_pureYM}
The YM gauge move is similar to the one generally employed 
in lattice gauge theories with flat background. 
After the random selection of a link $l$ of the dual lattice, 
we have to extract a new value of the link variable $U_l^\prime$ from the gauge group $G$. In order to do that, we have chosen a standard heat-bath algorithm,
meaning that the new link variable 
$U_l^\prime$ is selected according to the distribution fixed by the 
heat bath of nearby gauge link variables.

\begin{figure}
\centering
\includegraphics[width=0.5\textwidth]{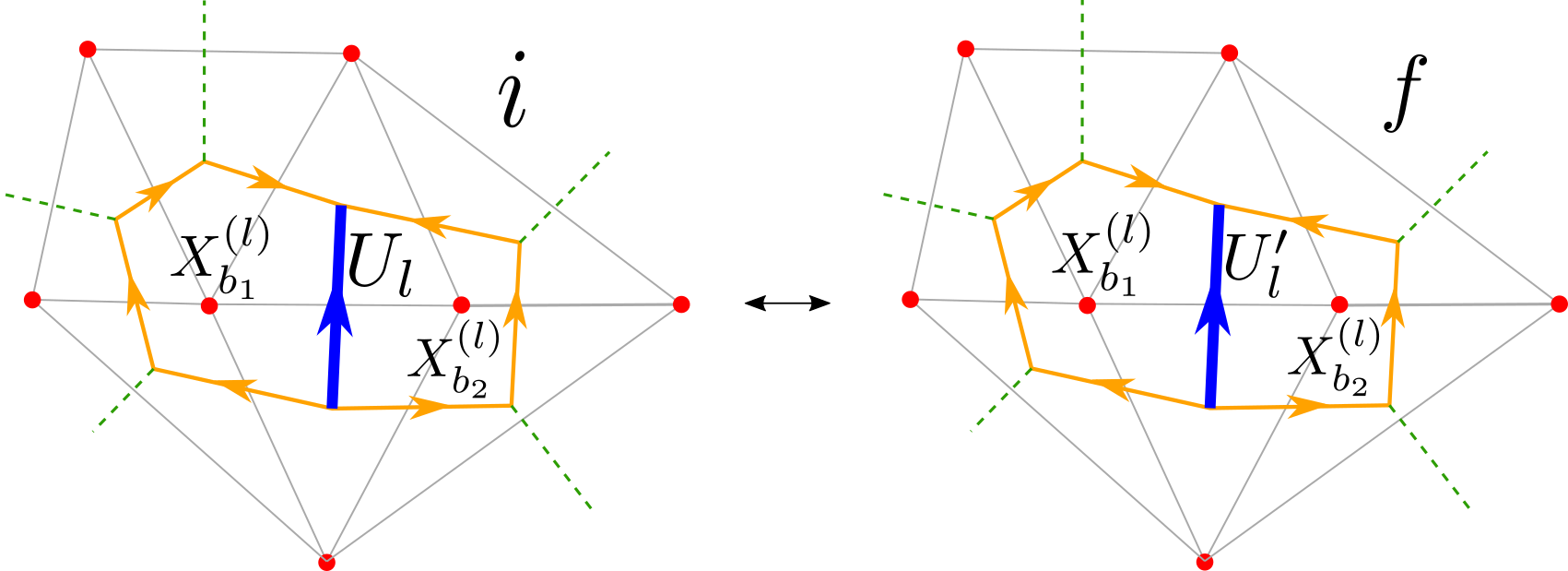}
\caption{Sketch of a typical gauge move, where the only difference between the initial 
    and final state is the value of the link variable associated to the edge $l$ (in blue)
    of the graph dual to the triangulation. 
    The orientation of the link products for the staples 
$X^{(l)}_{b_1}$ and $X^{(l)}_{b_2}$ is indicated by the arrows (in the orange path).}
\label{fig:movegauge_draw}
\end{figure}

In practice, such distribution can be expressed
in terms of the local \emph{force}
$F_l \equiv \sum\limits_{b \ni l} (X_b^{(l)}/n_b)$
acting on that link, where 
$X_b^{(l)}$ is the \emph{staple} 
going around the bone $b$,
i.e., the ordered product of the 
other links that form the plaquettes around $b$ that contains $l$, as 
sketched in Fig.~\ref{fig:movegauge_draw}.
The distribution stemming from Eq.~(\ref{eq:un-YMact_cdt}) is
\beq
P(U_l^\prime ) d U_l^\prime  \propto 
d U_l^\prime \exp \left( \frac{\beta}{N} \textrm{Re Tr} 
\left[ U_l^{\prime} F_l^{\dagger} \right] \right) .
\eeq
where $d U_l^\prime$ stands for the Haar measure over the gauge group.
The distribution above is sampled using standard algorithms
already employed in $U(1)$ or $SU(N)$ lattice gauge theories~\cite{hb_creutz,hb_kennedy-pendleton}.

\subsubsection{Triangulation move $(2,2)$}
\begin{figure}
\centering
\includegraphics[width=0.5\textwidth]{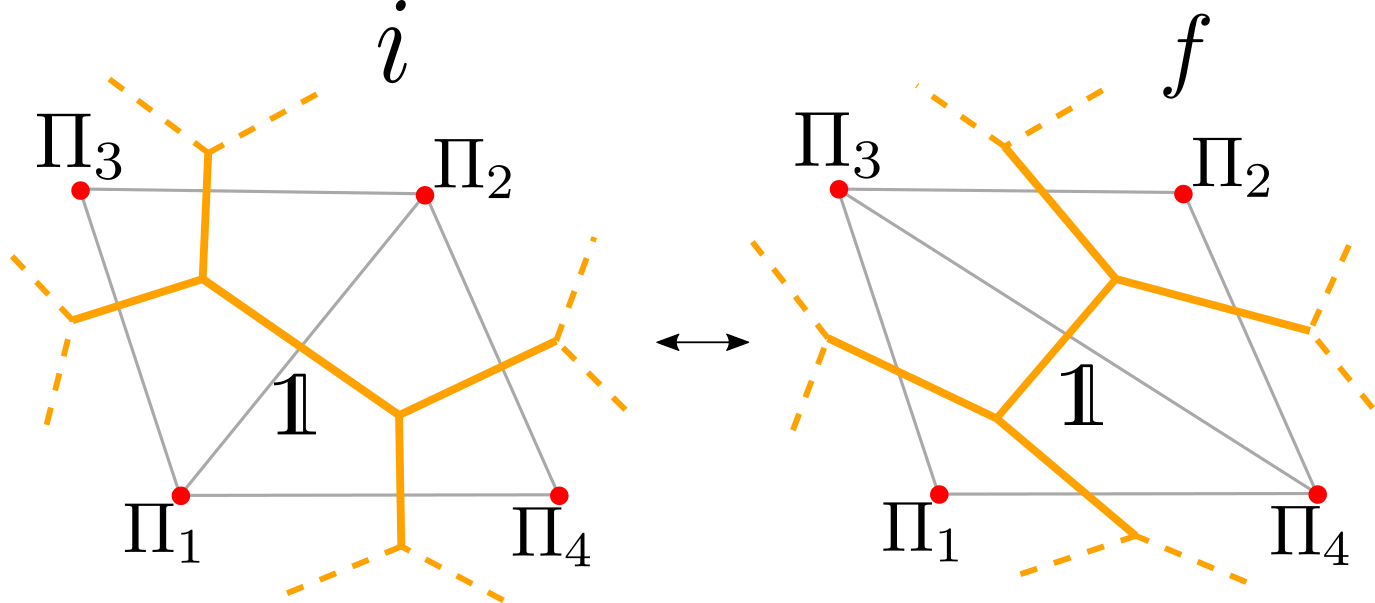}
\caption{Sketch of a $(2,2)$ move, where the edge dual to the flipped link is gauge fixed to the identity. Notice that, even if the plaquettes involved $\Pi_{1\leq i \leq 4}$ keep the same value after the move, their contribution to the action does change because the coordination numbers of all the four involved 
vertices change.}
\label{fig:move22_draw}
\end{figure}

As shown in Fig.~\ref{fig:move22_draw}, the move 
$(2,2)$ consists in flipping a time-like edge of the triangulation
and thus the associated dual link. Under such a move, the
volume of the triangulation does not change and the purely
gravitational part of the action remains constant.
The change of the triangulation also changes the space $\YMconfs$ 
of the gauge configurations, however, because of the gauge freedom,
this effect can be easily accounted for.
\label{subsubsec:un-db_move22}

Indeed, the links are not gauge invariant quantities, so in every
node of the dual graph we can always perform a gauge transformation
as the action of an element $g \in G$ of the group on the links
that enter and exit the node.
In particular, in both the initial and the final configurations
we can always set the value of the central link equal to the 
group identity $\mathds{1} \in G$. 
That permits to easily identify the final gauge
configuration apart from an irrelevant gauge 
transformation, so that a Metropolis--Hastings 
accept-reject step can be performed, as described 
in more details in the Appendix.

\subsubsection{Triangulation $(2,4)$-$(4,2)$ moves}\label{subsubsec:un-db_move2442}
\begin{figure}
\centering
\includegraphics[width=0.5\textwidth]{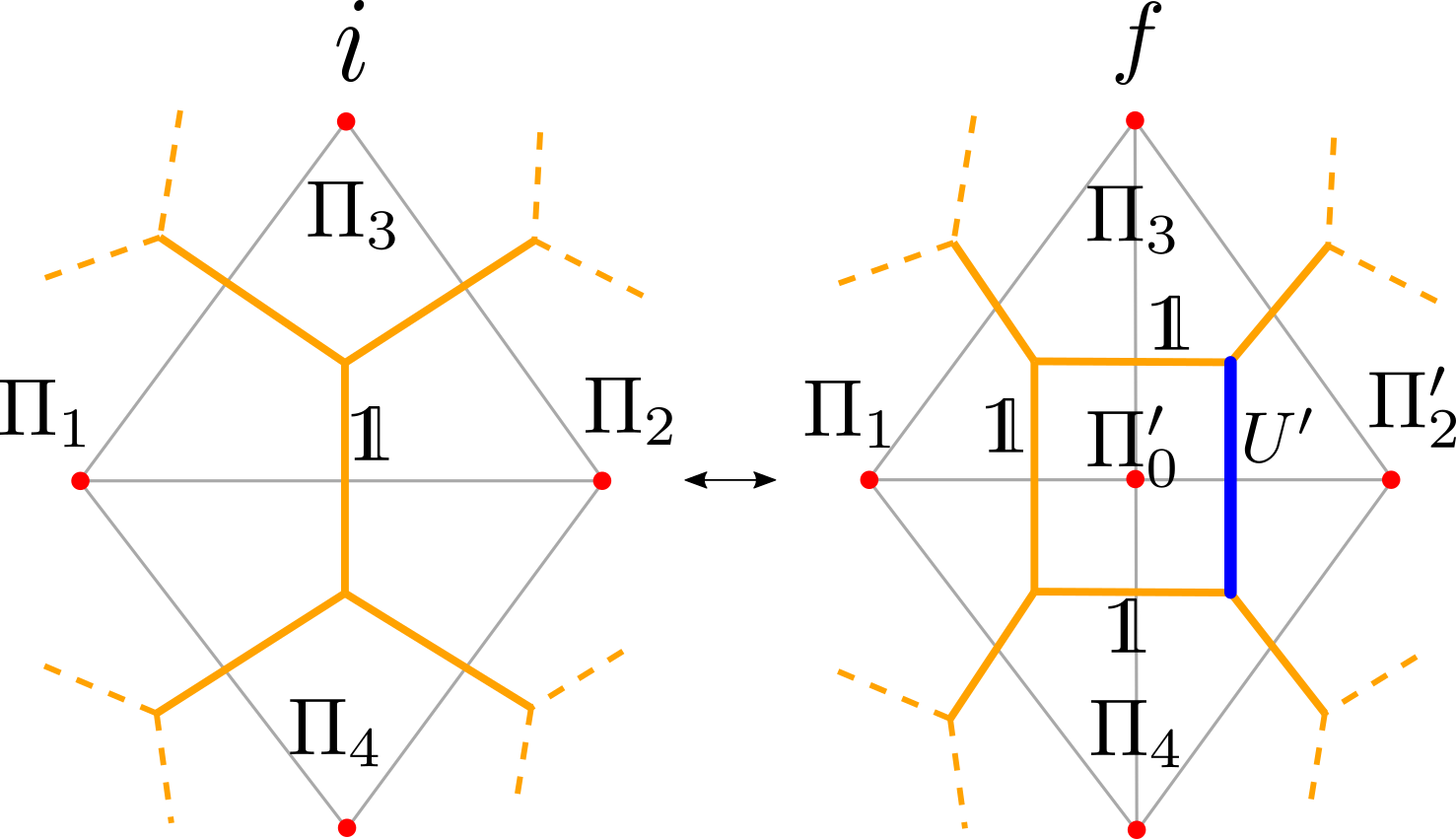}
\caption{Sketch of a $(2,4)$ move and its inverse. The link variable $U^\prime$ (in blue) is the newly extracted one, as described in more details 
in the Appendix.}
\label{fig:move2442_draw}
\end{figure}

Unlike the other moves, the $(2,4)$ and $(4,2)$ ones
change the volume of the triangulation, respectively creating
and eliminating two triangles and a vertex of coordination four,
giving a variation of the CDT part of the action equal to
$\Delta S^{(2,4)}_{CDT} = - \Delta S^{(4,2)}_{CDT} = 2 \lambda$.
The effect on the gauge configuration is thus, respectively,
the introduction or the removal of a central plaquette 
$\Pi_0^\prime$ of length four, as we can see from 
Fig.~\ref{fig:move2442_draw}. 

Since the plaquettes are gauge invariant observables, gauge 
transformations can set to the identity $\mathds{1}$ at most 
three out of four links of the central plaquette $\Pi_0^\prime$, shifting all 
the physical content in the remaining one.

Let us consider for example the $(2,4)$ move. 
When performing this move,
the plaquette that we introduce must in general have a non-trivial value. 
This is necessary in order for this move to be 
the inverse of the $(4,2)$ one, in which the value of the 
destroyed $\Pi_0^\prime$ plaquette is arbitrary and not fixed by us.
Therefore, the $(2,4)$ move must involve the extraction of a new 
link variable $U^\prime$, which also changes the 
value of the plaquette $\Pi_2^\prime$. Details on our choice 
for the extraction probability are reported in the Appendix.

The variation of the YM gauge action now receives a contribution both
from the variation of the coordination number $n_b$ 
for the preexisting plaquettes 
$\Pi_1$, $\Pi_3$ and $\Pi_4$ and from the creation of the link variable
$U^\prime$. The move is completed by a Metropolis--Hastings accept-reject step.

\section{Numerical Results}\label{sec:numres}

In this Section we present and discuss the results of our numerical simulations
of 2D CDT minimally coupled to Yang-Mills theories, considering either
$G=U(1)$ or $G=SU(2)$ as a gauge group. We briefly recap the bare 
parameters entering the discretized theory: the cosmological parameter
$\lambda$, defined in Eq.~(\ref{eq:mincoup-CDT2Daction}), 
and the inverse gauge coupling $\beta \propto 1/g^2$, defined in 
Eq.~(\ref{eq:un-YMact_cdt}). We notice that the definition of 
$\beta$ differs from other definitions reported in the 
literature for discretized Yang-Mills theories by a proportionality 
factor, due to the explicit presence of the 
bone coordination numbers $n_b$ 
in the expression of the action, but also due to the fact that 
the reference dual lattice in the flat limit is hexagonal.
In any case, nothing changes regarding the expected running of $\beta$
as the continuum limit of pure Yang-Mills is approached, 
i.e., $\beta \propto 1/a^2$ for a two-dimensional theory.

A further parameter characterizing the sampled triangulations is the 
number of temporal slices $N_t$, which is fixed, while the spatial extension
of each slice is a dynamical variable and is part of the observables
studied in the following. The value of $N_t$ has been fixed for each simulation
in order to guarantee that we dealt with negligible finite size effects: this is ensured
by comparing $N_t$ with the system correlation lengths in the temporal 
direction, either gravity related or gauge related, that we will define and 
discuss 
in the following.
The overall imposed topology is toroidal, 
i.e., with periodic boundary conditions both in the temporal and spatial 
directions: that amounts, because of the Gauss-Bonnet theorem in 
two-dimensions, to globally flat geometries, meaning that
we expect $\langle n_b \rangle = 6$.
 
In the following we start by describing gravity related observables, which are 
then analyzed to sketch the phase diagram of the discretized models in the 
$\lambda - \beta$ plane and compare it with general expectations based 
 on a strong coupling (i.e., small $\beta$) expansion. Then we illustrate
the behavior of some gauge observables,
in particular we consider the correlation function of the holonomies (torelons)
and
for the $U(1)$ case, the analysis of topological observables. 
The latter will give us illuminating hints on how a variable
geometry can strongly ameliorate well known algorithmic problems like
topological freezing.

\subsection{Gravity related observables}

We have considered the total volume of the triangulation $V \equiv N_2$ 
(i.e., the total number of triangles)
and the volume profiles, 
i.e., the number of 
spatial links (spatial edges of the triangles) 
$V(t)\equiv N_{1s}(t)$ in each slice, 
where $t$ labels the slice time. 
From 2d constraints, $V$ equals two times the integral of $V(t)$ and is expected 
to diverge at the phase transition; $V(t)$ contains more information,
in particular its two-point function is expected to show 
a critical behavior at the phase transition, with a diverging 
correlation length.

The connected two-point correlation function 
of $V(t)$ is defined as 
\begin{equation}\label{eq:un-corr2prof}
	C_{\text{Vprof.}}(\Delta t) \equiv \frac{\langle V(t) V(t+\Delta t)\rangle - \langle V(t)\rangle \langle V(t + \Delta t) \rangle}{\langle V(t)^2\rangle - \langle V(t)\rangle^2},
\end{equation}
with periodic conditions implied; in the actual computation we assumed
$\langle V(t + \Delta t) \rangle = \langle V(t) \rangle$ and 
took an average over all slices\footnote{This is meaningful
because we did not observe any breaking of the time-translation symmetry, 
which is observed instead in the $C_{dS}$ phase of 4D CDT.}.
In practice, we performed a blocked resampling of the expression 
in Eq.~\eqref{eq:un-corr2prof} over many configurations,
and we fitted the curve $C_{\text{Vprof.}}$ as a function of 
$\Delta t$ 
to an exponential function 
$\propto \exp(-\Delta t/\xi_{\text{Vprof}})$, 
in order to extract the \emph{correlation length of volume profiles},
 $\xi_{\text{Vprof}}$.

\begin{figure}
\centering
\includegraphics[width=0.5\textwidth]{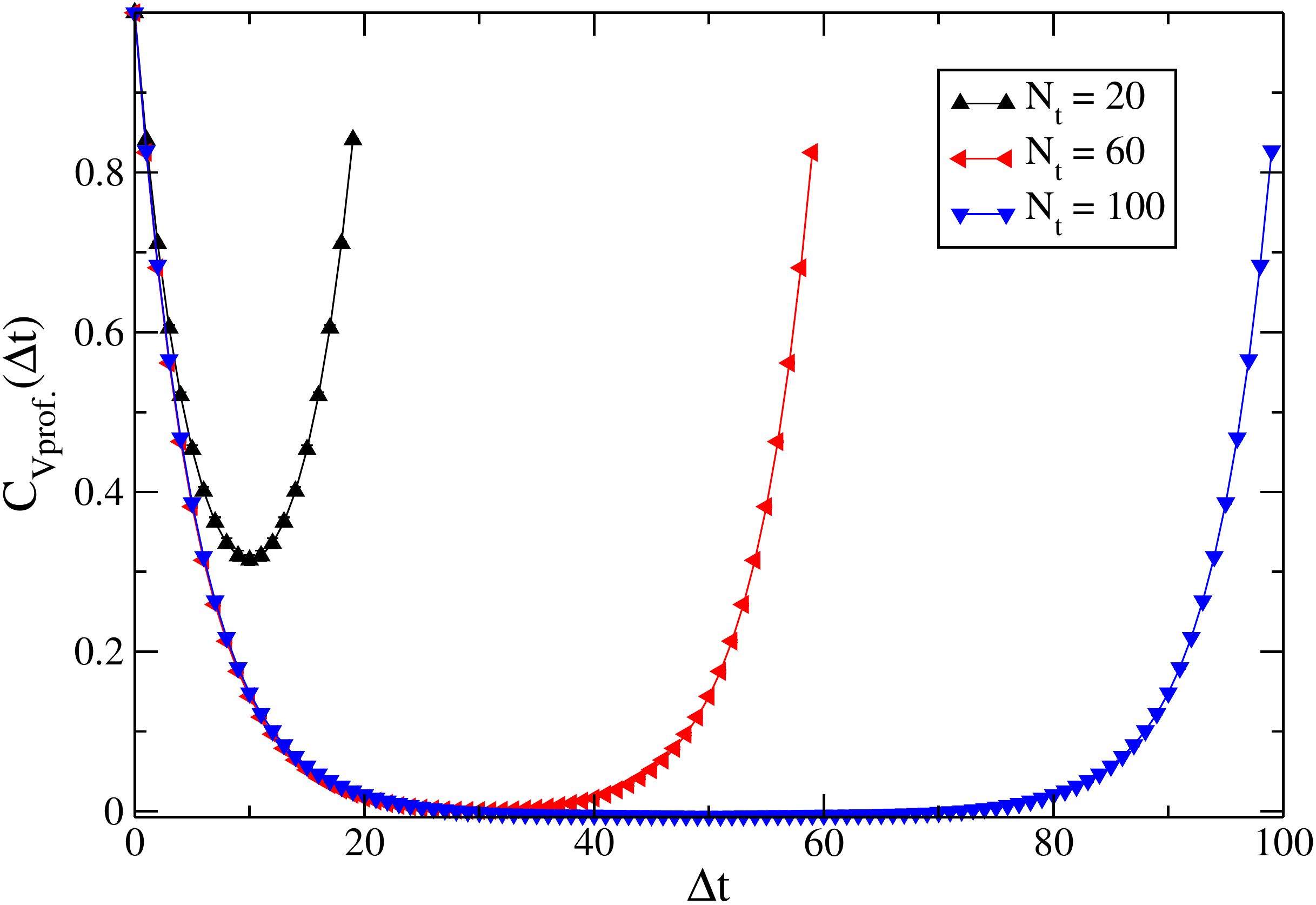}
\caption{$2$-point connected correlation function for the volume profiles 
at $\beta=0.01$, $\lambda=0.695871$ and three different 
number of slices: $N_t = 20, 60$ and $100$.}
\label{fig:corr2_Nt_20_60_100}
\end{figure}

Fig.~\ref{fig:corr2_Nt_20_60_100} shows
an example of the determination of the two-point 
function for $\beta=0.01$ and $\lambda=0.695871$
in the $U(1)$ case. 
We report determinations for three different values of $N_t = 20, 60$ and 100,
the correlation function is symmetric under half-lattice reflections 
by construction, because of the periodic boundary conditions.
Finite size effects are under control in this case
at least for the two largest values of $N_t$. Indeed,
an exponential fit limited to the first half 
of the lattice returns $\xi_{\text{Vprof}} = 5.20(2)$ for $N_t = 60$
and $\xi_{\text{Vprof}} = 5.26(2)$ for $N_t = 100$, the reported errors
include systematics related to the variation of the fit range, the 
reduced $\chi^2 / {\rm dof}$ test was of $O(1)$ in all cases.
A similar analysis has been performed for other simulation points.

\subsection{Phase diagram}\label{subsec:phase_diag}

{In two dimensions, the phase diagram of pure gravity CDT 
    presents a critical point for $\lambda_c=\log(2)\simeq 0.693147$ 
    (see Refs.~\cite{two_dim_scaling,two_dim_scaling2,cdt_review12}).
 Triangulations are weighted in the path-integral by 
 a $\exp(- \lambda V)$ factor, so that 
 the parameter $\lambda$ plays a role similar a chemical 
  potential coupled do $V$: lowering the value 
of $\lambda$ results in an increase of $\langle V \rangle$,
until it diverges as $\lambda \to \lambda_c$ from above.
In particular, we expect the average volume to present 
a critical behavior which can be fitted to a power law scaling of the form:
\begin{equation}\label{eq:un-scalingfunc}
    f(\lambda) = \frac{A}{(\lambda - \lambda_c)^\mu},
\end{equation}
where $\mu$ is a critical index.
The correlation length $\xi_{V_{prof}}$ is expected
to follow a similar behavior, although 
with a different critical index $\nu$.
}

The effect of the introduction of the gauge coupling
can be estimated  
in the strong coupling (low $\beta$) limit. 
 Indeed, for $\beta \to 0$ gauge links are randomly 
 distributed and the plaquette averages to zero. Therefore, 
taking into account that by the 
Gauss-Bonnet theorem triangulations have zero
average curvature, i.e.~$\langle n_b \rangle = 6$, 
the typical action of a gauge configuration
is expected to be approximately
$\beta N_0 /6 = \beta N_2 /12$.
That can be seen as an effective shift
of the $\lambda$ parameter to
$\widetilde{\lambda} \simeq (\lambda+\beta/12)$,
so that the critical threshold 
is expected to move to 
$\widetilde{\lambda}_c = \log(2) \implies \lambda_c^{(G)}(\beta) =  \log(2) - \beta/12$.

\begin{figure}
\centering
\includegraphics[width=0.5\textwidth]{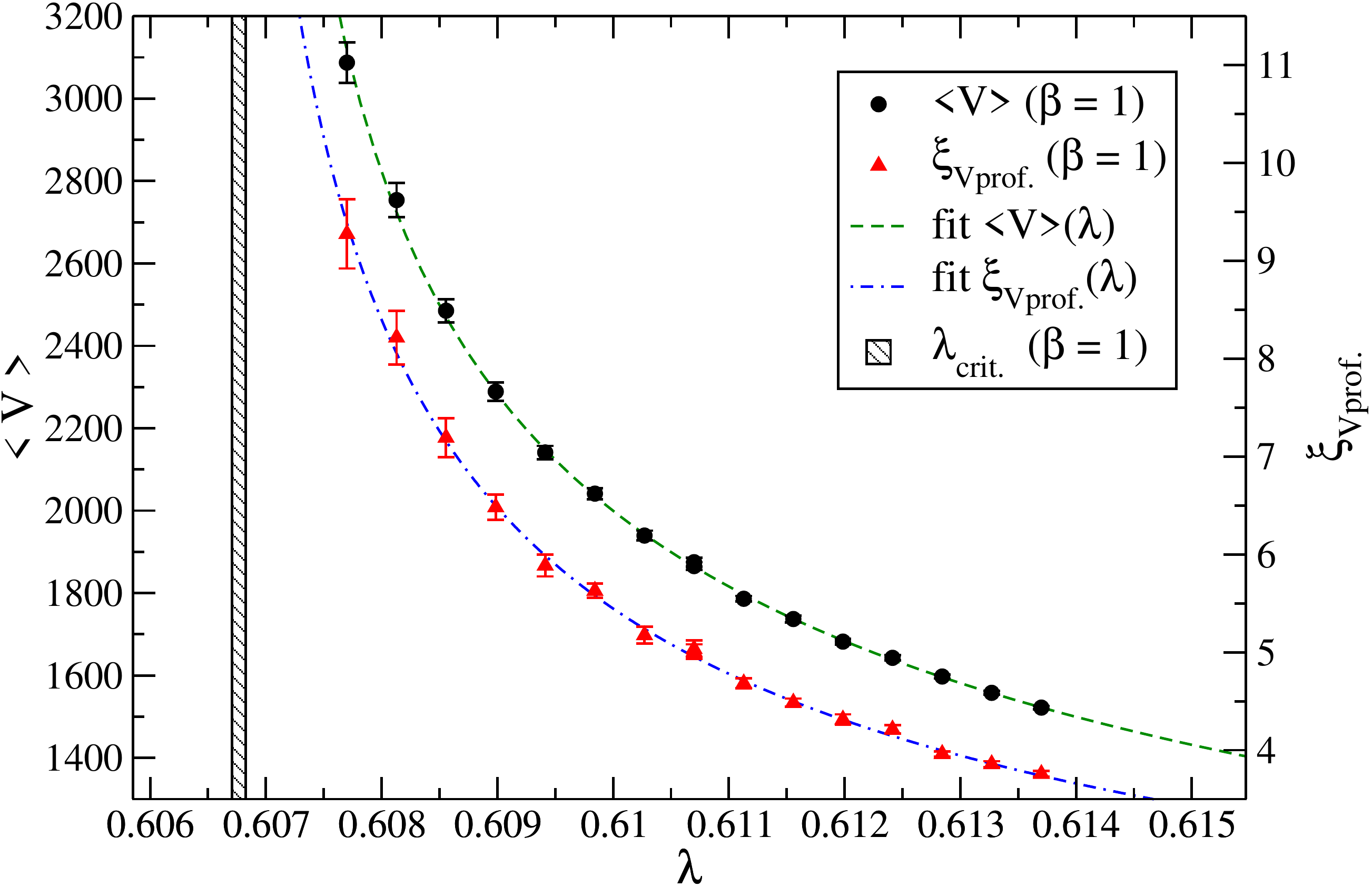}
\caption{Determination of $\langle V \rangle$ and 
$\xi_{\text{Vprof.}}$ as a function of $\lambda$ for $U(1)$
at $\beta = 1$. We report the results of a fit to a power law behavior
as in Eq.~(\ref{eq:un-scalingfunc}) (see text).}
\label{fig:u1critfit_beta1_multiple}
\end{figure}

In Fig.~\ref{fig:u1critfit_beta1_multiple} we show results
obtained for the average volume $\langle V \rangle$ and for the 
correlation length from simulations with gauge group $U(1)$ 
at $\beta = 1$. A fit according to Eq.~\eqref{eq:un-scalingfunc}
works perfectly for both $\langle V \rangle$ and the 
correlation length, obtaining 
$\lambda_c^{(G)} =  0.6068(1),\ \mu = 0.357(7)$ for the former
($\chi^2/{\rm dof} = 5.7/13$) and 
$\lambda_c^{(G)} =  0.6064(2), \ \nu = 0.53(3)$ for the latter
($\chi^2/{\rm dof} = 8.6/13$): that confirms that 
the two quantities lead to a consistent determination
of $\lambda_c$ (even if with a different critical index)
and that the critical value is shifted below $\log 2$.

\begin{figure}
\centering
\includegraphics[width=0.5\textwidth]{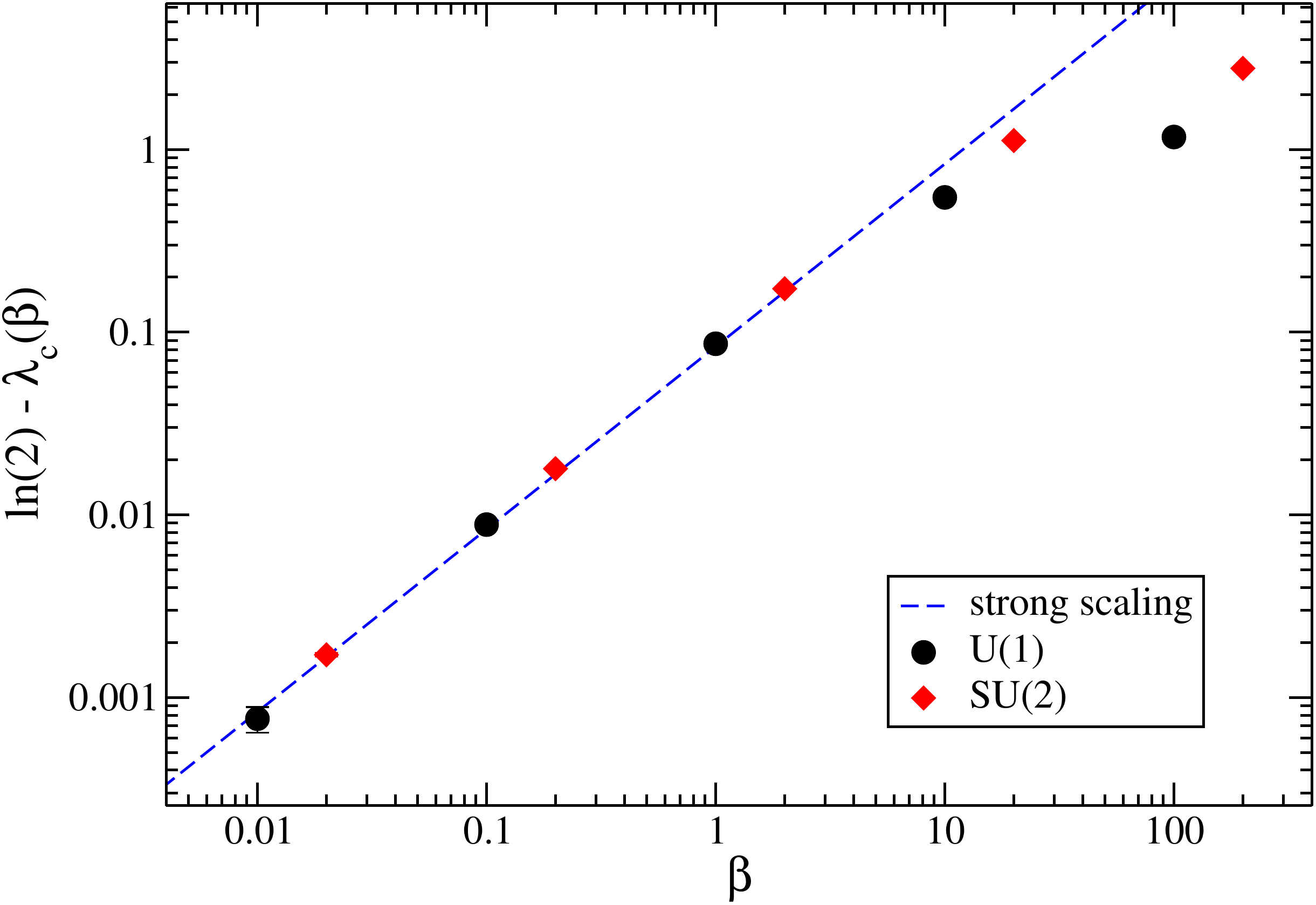}
\caption{Critical points for the $U(1)$ and $SU(2)$ gauge theories minimally coupled to 
    two-dimensional CDT as a function of $\beta$.
}
\label{fig:pseudocritical}
\end{figure}

\begin{figure}
\centering
\includegraphics[width=0.5\textwidth]{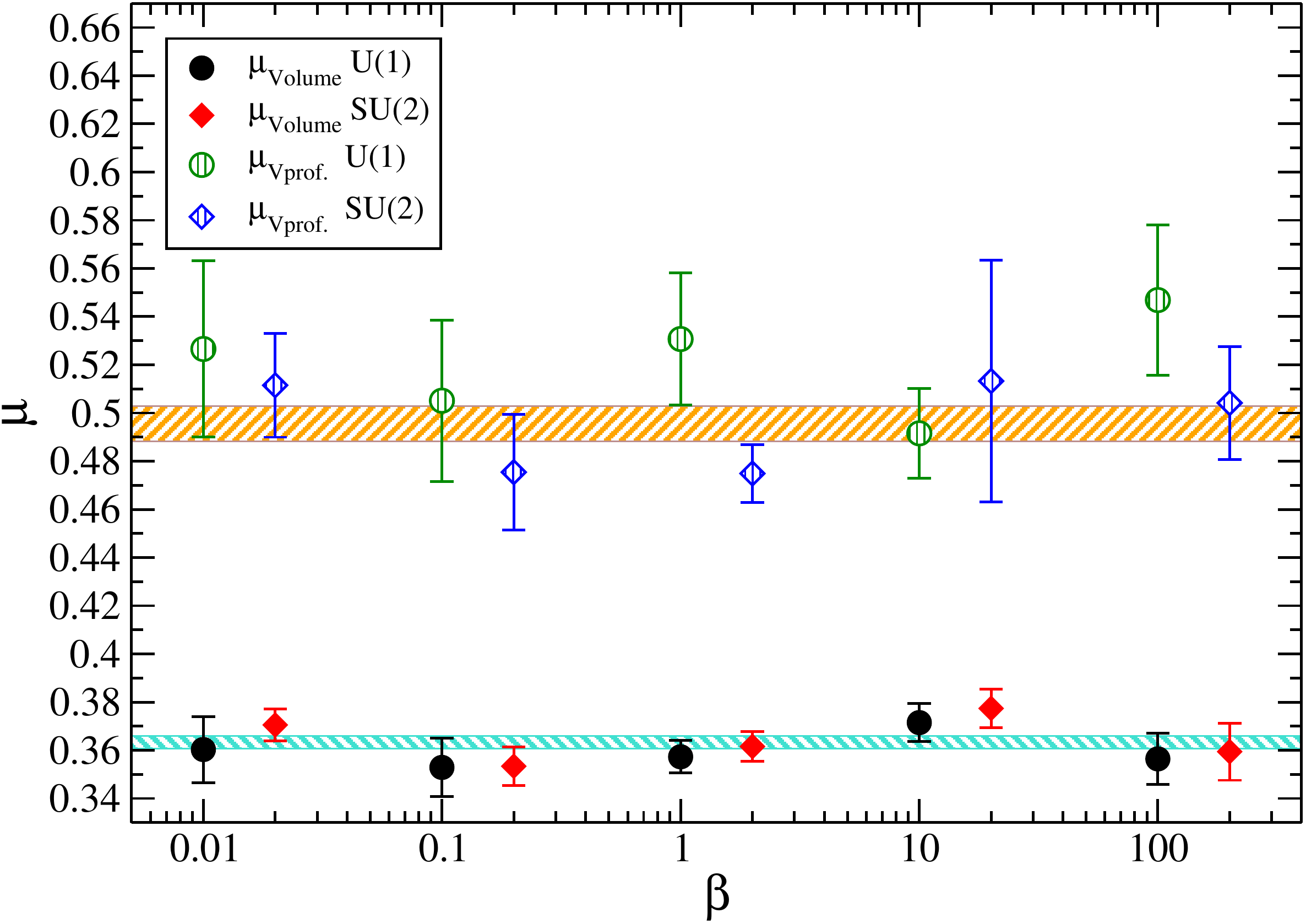}
\caption{Critical indices $\mu$ and $\nu$
associated respectively
to $\langle V \rangle$ 
and $\xi_{\text{Vprof.}}$ (see Eq.~\eqref{eq:un-scalingfunc}), plotted  
as a function of the bare gauge coupling
$\beta$. Data for the gauge groups $G=U(1)$ and $SU(2)$
have been slightly shifted horizontally to improve readability.
The horizontal bands represent the results
of a fit to constant values discussed in the text.}
\label{fig:fit_critidx}
\end{figure}

A similar analysis has been repeated for different values of $\beta$,
both for $U(1)$ and $SU(2)$.
In Fig.~\ref{fig:pseudocritical} we report 
the quantity $\log 2 - \lambda_c^{(G)}(\beta)$ obtained 
in the two cases and over a wide range, comparing it with the strong 
coupling prediction discussed above, which seems to work perfectly 
until $\beta \lesssim 1$. What is more interesting is to look 
at the critical indices, which are reported in 
Fig.~\ref{fig:fit_critidx} and appear to be independent 
of $\beta$ in the whole range, suggesting that the only effects of the 
gauge fields on gravity observables consist in a shift of 
$\lambda_c$.
On the basis of the apparent stability of the critical indices,
we have performed a fit of all determinations to a constant
value, obtaining $\mu = 0.363(3)$ ($\chi^2 / \textrm{dof} = 9/9$)
and $\nu = 0.496(7)$ ($\chi^2 / \textrm{dof} = 9.1/9$).

One could argue that this is not unreasonable, since
 it is known (see, e.g., the discussion 
in Refs.~\cite{cdtgauge_anal,Cao:2013na,bonati_flatsuscu1}) that  
two-dimensional gauge fields can be easily integrated away:
 after a proper gauge fixing which freezes spatial dual gauge 
 links, one is left with a theory of plaquettes and the path 
 integral over them can be performed.
 Nevertheless, on a curved geometry one is left
 with a non-trivial contribution to the remaining path-integral, 
  depending on the coordination
 numbers $n_b$: according to our numerical results, such contribution
 seems to be irrelevant, at least for what concerns the critical properties 
 of the system. It is interesting to notice that the critical 
 index of the correlation length turns out to be compatible
 with $1/2$ within errors, i.e.~with a mean field behavior which 
 could explain why the local fluctuations of $n_b$ are 
 not relevant.

%

\subsection{Gauge observables}\label{subsec:gauge_obs}

\begin{figure}
\centering
\includegraphics[width=0.5\textwidth]{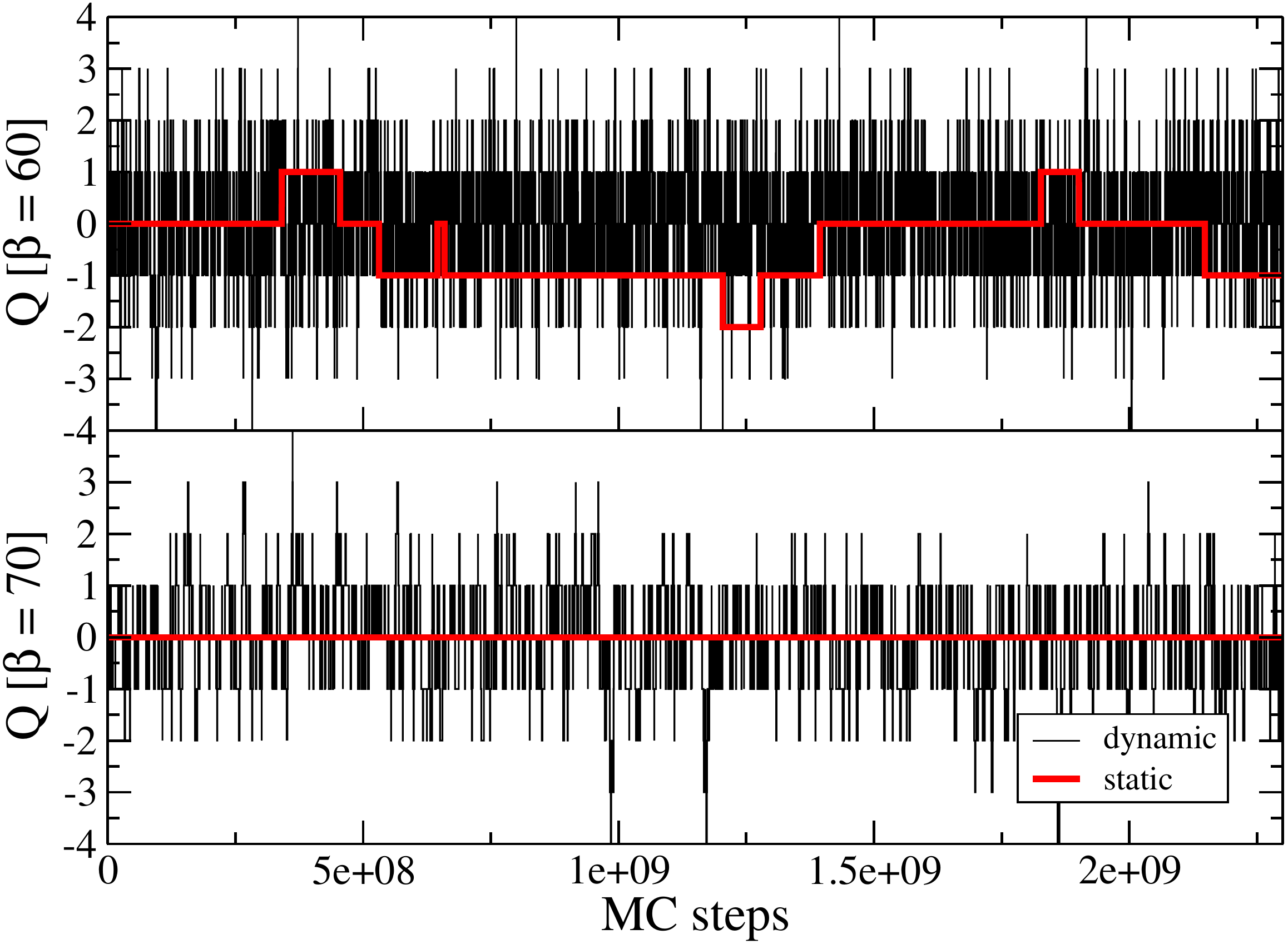}
\caption{Comparison of topological charge histories at $\beta=60$ and $70$
    for the static (flat) and dynamic simulations at average volume $N_2=800$.}
\label{fig:Qhistories_compare}
\end{figure}

The first gauge observable considered in our study 
is the topological charge, or winding number of the gauge field.
It counts the winding of the gauge field at infinity and, in two-dimensions,
is non-trivial only for the $U(1)$ gauge group: in this case it amounts
to the total flux through the two-dimensional manifold of the 
field strength, which is quantized for a compact manifold like a torus
or if vanishing conditions are imposed for the field strength at infinity.
A possible discretization, keeping integer values as in the 
continuum, is the following:
\begin{equation}\label{eq:un-topcharge_U1}
	Q \equiv \frac{1}{2 \pi}\sum_{b\in \mcTau^{(d-2)}} \arg\lbrack Tr (\Pi_b)\rbrack,
\end{equation}
where $\Pi_b$ is the plaquette around the vertex $b$, 
and the argument produced by $\arg z$ is restricted in the interval $(-\pi,\pi\rbrack$.
The cumulants of the topological charge distribution are related to the 
expansion of the free energy density $f$ of the gauge 
theory in powers of the topological
parameter $\theta$. In particular the topological susceptibility,
$\chi \equiv \langle Q^2 \rangle /V$, fixes the leading order 
quadratic dependence, $f(\theta) = \chi \theta^2 /2 + O(\theta^4)$. 

The fact 
that gauge configurations relevant to the 
path-integral divide, in the continuum, in homotopy classes 
characterized by different integer values of $Q$, is at the origin of 
the infamous critical slowing down of topological modes in Monte Carlo
simulations, also known as {\em topological freezing}, 
when the continuum limit is approached~\cite{Alles:1996vn,top_freeze,Luscher:2011kk,Bonati:2017woi}: the tunneling between
different topological sectors becomes more and more suppressed, as 
it involves the sampling of large action configurations.
It is then interesting to understand what is the impact of an underlying 
fluctuating geometry on this problem.

There are at least two reasons to expect such an impact, and to expect
that it is in the 
direction of an improvement:
\emph{i)} the creation/destruction
of portions of space-time and of the gauge fields living on them
may in principle enhance the tunneling events; \emph{ii)}
plaquettes enter the action in Eq.~(\ref{eq:un-YMact_cdt})
divided by a $n_b$ factor, which however is not present in the topological 
charge in Eq.~(\ref{eq:un-topcharge_U1}), so that plaquettes centered 
around bones with a high coordination number $n_b$ are possible sources
of large local fluctuations of the field strength flux, with a 
special discount in the action.

\begin{figure}
\centering
\includegraphics[width=0.5\textwidth]{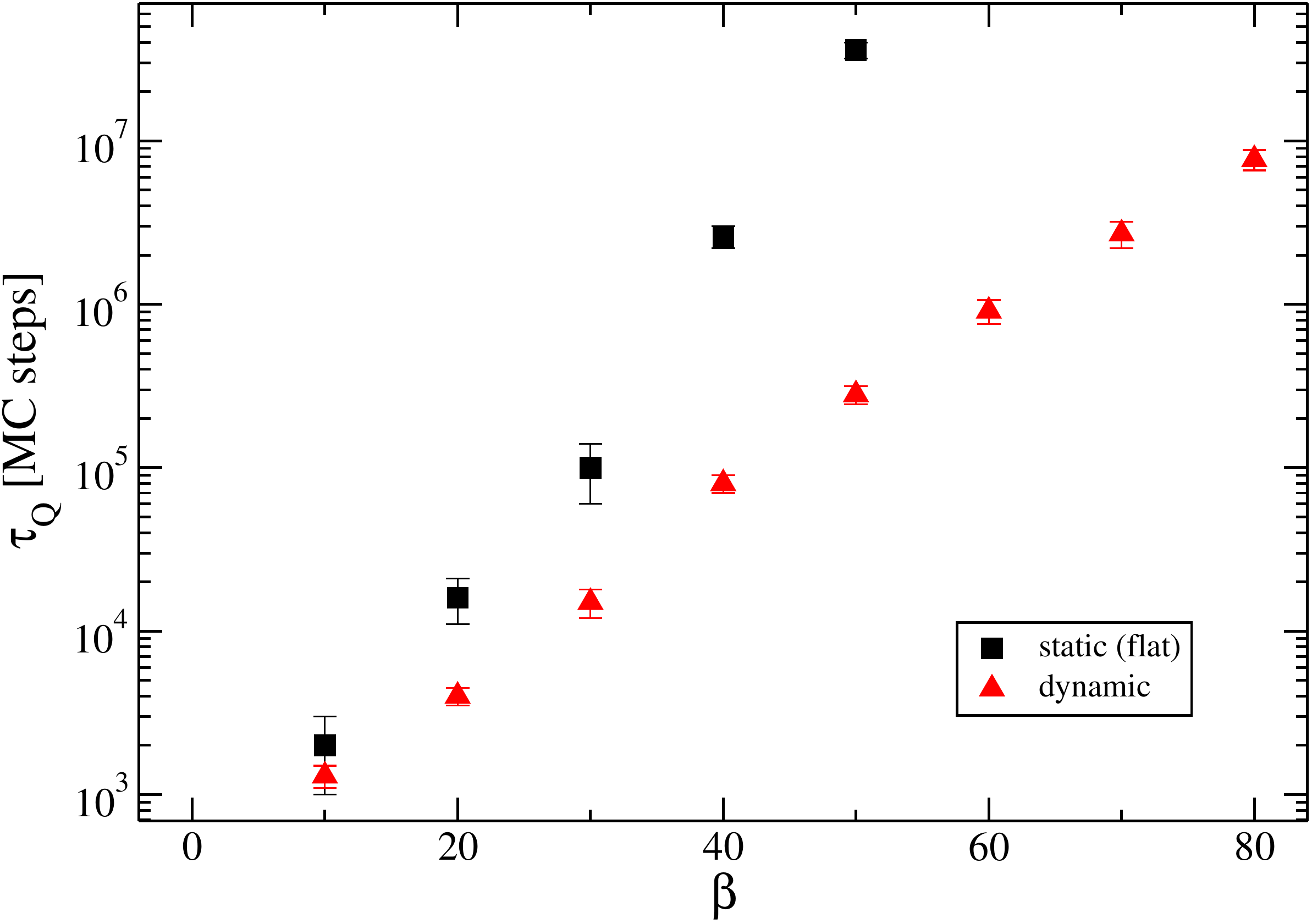}
\caption{Comparison in topological charge autocorrelation times 
    for the static (flat) and dynamic simulations with gauge group $U(1)$ 
    at average volume $\langle N_2 \rangle =800$.
    Measures with higher autocorrelation times than the ones shown have been dropped due to 
    not enough statistics or to algorithmic freezing.}
\label{fig:autocorrs_compare_20x20}
\end{figure}

A strong improvement in topological freezing is indeed what we observe.
For the sake of comparison, we have considered also simulations of
the compact $U(1)$ gauge theory 
on a static toroidal (i.e., flat) hexagonal 
lattice\footnote{This has been realized simply by initializing the triangulation to a torus 
of given sizes, 
and performing as Monte Carlo steps only gauge moves, without triangulation moves that 
would change the lattice geometry.}. 
Fig.~\ref{fig:Qhistories_compare} shows a comparison between 
the Monte Carlo histories of the
topological charge for the static and dynamical
case and for two values of $\beta$, $\beta=60$ and $70$;
the comparison has been made for an equivalent total volume,
namely $V=800$ for the static flat lattice
and $\lambda$ tuned so as to fix $\langle V \rangle \simeq 800$ for the dynamical 
one, with $N_t = 20$ in both cases.
The emergence of topological freezing is quite clear in static simulations
and $Q$ is completely frozen for $\beta = 70$:
on the contrary, 
this is not observed in dynamical simulations and
not even for higher values of $\beta$.
Autocorrelation times of $Q$ are shown 
as a function of $\beta$
in Fig.~\ref{fig:autocorrs_compare_20x20}:
the increase is at least exponential in the static case and
becomes prohibitive for $\beta \gtrsim 50$, while it is orders 
of magnitudes smaller and at most exponential
in the dynamical case.

\begin{figure}
\centering
\includegraphics[width=0.5\textwidth]{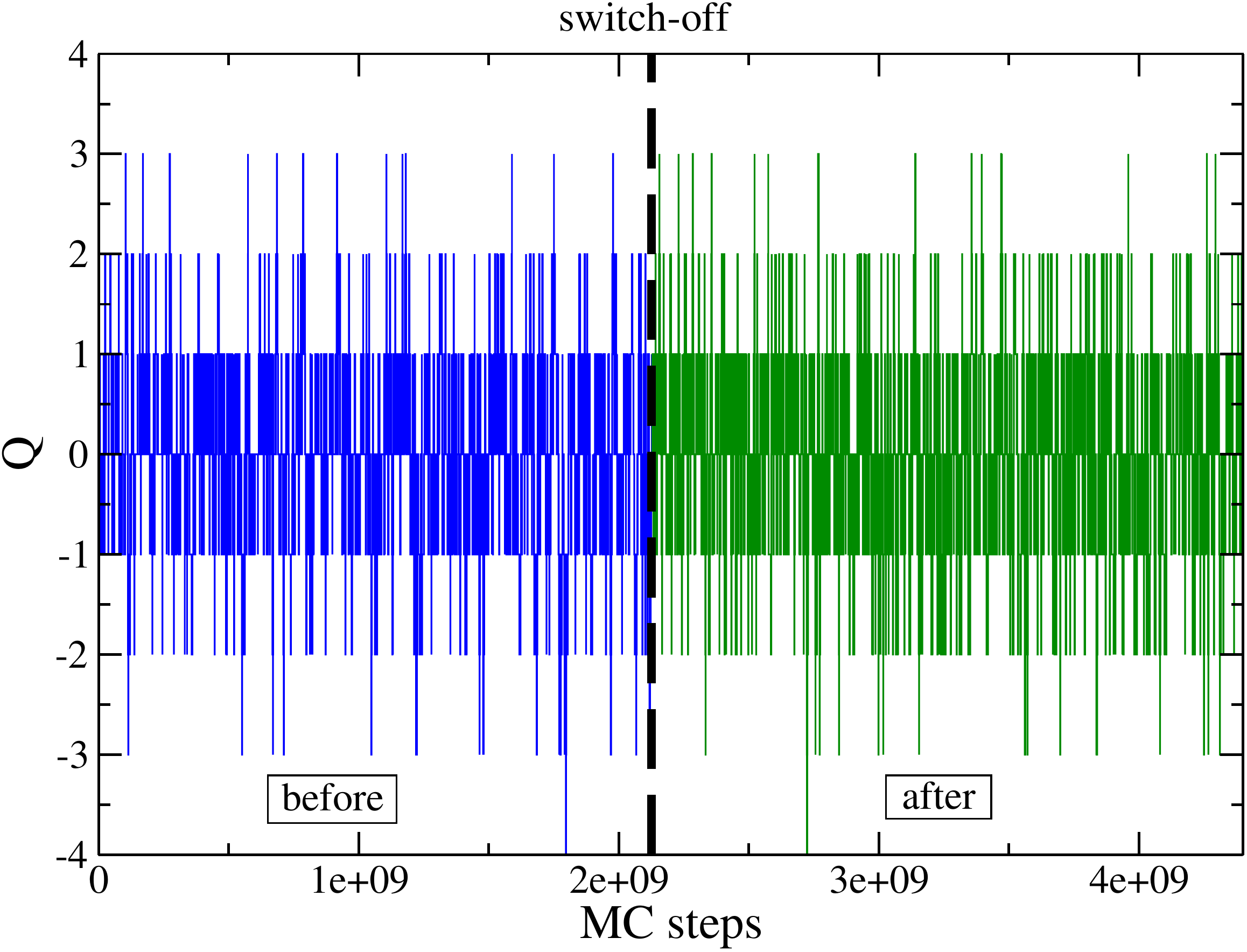}
\caption{History of the topological charge for a simulation with gauge group $U(1)$ 
    coupled to (dynamical) two-dimensional CDT at $\beta=70$
    where we switched off all geometry updates at a certain Monte Carlo step (dashed line), 
    after which the volume does not change ($V=800$).
}
\label{fig:Qhistory_staticurved}
\end{figure}

In order to better understand the origin of such improvement 
and differentiate between the two possible hypotheses exposed 
above, we have performed the following experiment. We have considered
the dynamical simulation at $\beta = 70$ and we have stopped  
updating the geometry from a certain step on, i.e., we have 
continued the Markov chain performing only pure gauge moves on a fixed, 
but non-flat,
triangulation. In this way the possible improvement coming from
a non-constant geometry stops, while the possible improvement coming 
from a non-uniform geometry, i.e.~from 
a space-time lattice with a locally variable curvature, remains.
The result of the experiment is shown in Fig.~\ref{fig:Qhistory_staticurved},
which clearly demonstrates that most of the improvement
comes from the non-flatness of the geometry. From an intuitive perspective, 
vertices with a large negative curvature (large $n_b$) represents
large holes in the lattice where fluctuations can happen more easily 
leading to tunneling events between topological sectors.
Such a dramatic improvement in the autocorrelation times of $Q$
could be useful also in standard simulations of lattice gauge theories,
where temporary fluctuations of the underlying geometry could 
be considered as a way to speed up the updating of topological modes.

\begin{figure}
\centering
\includegraphics[width=0.5\textwidth]{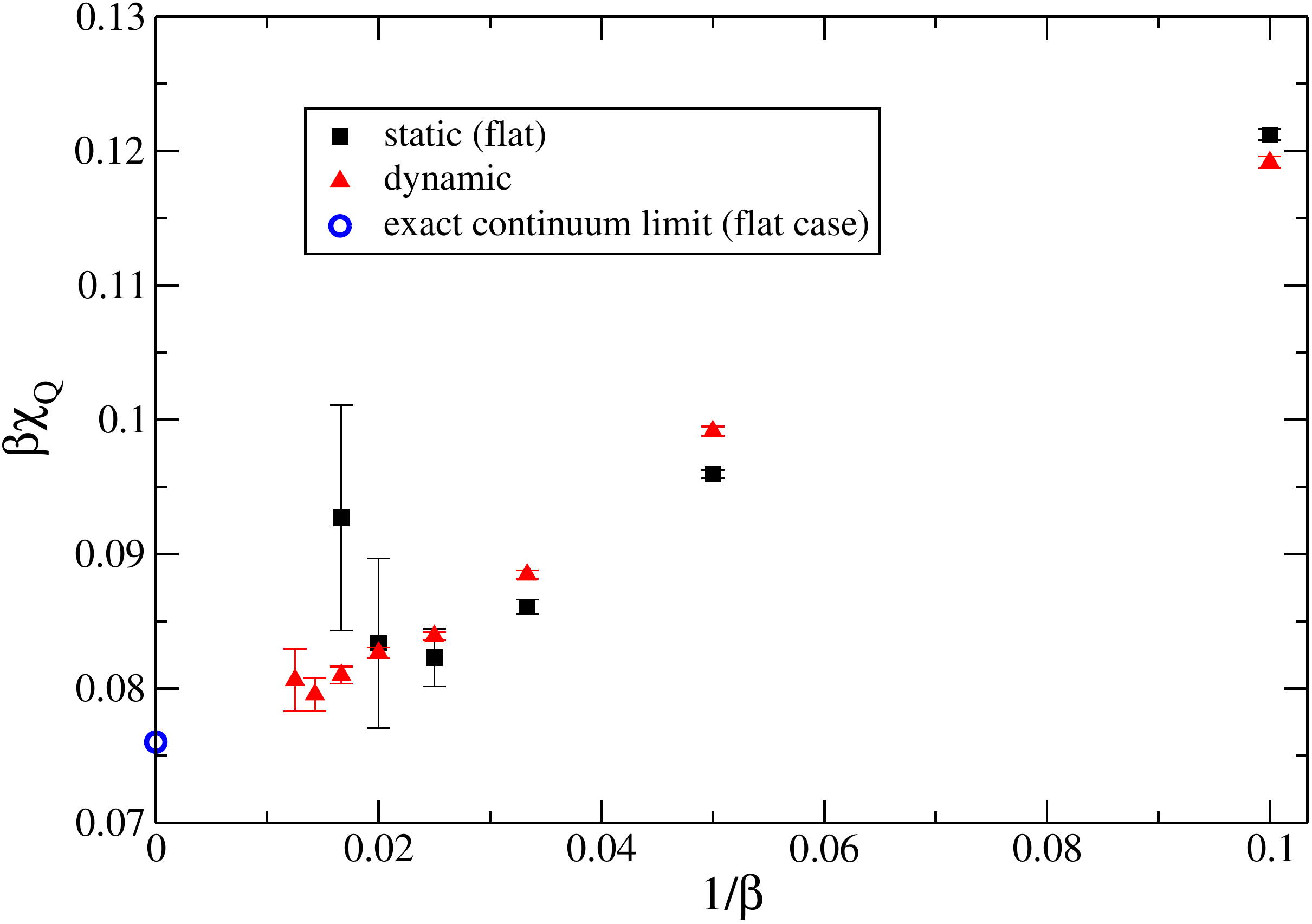}
\caption{Comparison in topological susceptibilities for the static (flat) and dynamic 
simulations at average volume $N_2=800$.}
\label{fig:topsusc_stat_dyn_20x20}
\end{figure}

Fig.~\ref{fig:topsusc_stat_dyn_20x20} shows a comparison between the topological susceptibilities 
(times $\beta$) measured from the same simulations
for the  flat static and dynamical case; in the dynamical case
the susceptibility has been defined as 
$\chi_Q \equiv \langle \frac{Q^2}{V}\rangle$,
i.e., the volume fluctuations have been taken into 
account as well,
given that, unlike the static case, 
$\langle V \rangle = 800$ only on average
in this case\footnote{Other definitions of susceptibility can ben considered, 
for example $\frac{\langle Q^2 \rangle}{\langle V \rangle}$ 
or $\langle (Q/V)^2 \rangle \langle V \rangle$, but the differences observed are negligible,
pointing out that $Q$ and $V$ do not exhibit a strong correlation.}.
The susceptibilities determined in the dynamical case slightly differ from 
those obtained in the static case, pointing out to an influence of the gravity 
coupling on gauge observables. However, in the dynamical case it is possible 
to obtain precise data at relatively large values of $\beta$, thus 
permitting a sort of (wrong) continuum limit at fixed bare gravity coupling,
which, assuming $1/\beta$ corrections and considering only data
with $\beta > 30$, turns out to be $\beta \chi_Q = 0.0758(14)$ 
with a reduced chi-squared $\chi^2/\textrm{dof} = 1.6/3$: 
this is in agreement with the analytical 
prediction for the flat continuum theory~\cite{bonati_flatsuscu1}, which, taking into
account the additional factor 6 in our definition of $\beta$, is $\beta \chi_Q = 3 / (4 \pi^2)$.
\\

What we have considered above has been correctly defined as a  
\emph{wrong continuum limit}. Indeed, in order to obtain a 
\emph{correct continuum limit} one should ensure that all correlation lengths,
both gravity and gauge related, diverge; 
in the previous analysis, instead, we have considered either 
the $\lambda \to \lambda_c$ limit at fixed $\beta$, or 
the $\beta \to \infty$ limit at fixed average volume.
A proper continuum limit will be obtained moving along particular 
curves $\lambda(\beta)$ in the $\lambda-\beta$ plane, 
the so-called \emph{lines of constant physics},
along which the ratios of different correlation lengths is kept 
fixed while all of them diverge as $\beta \to \infty$. Each
line of constant physics, corresponding 
to different ratios of correlation lengths, 
will define a different renormalized quantum theory.

\begin{figure}
\centering
\includegraphics[width=0.5\textwidth]{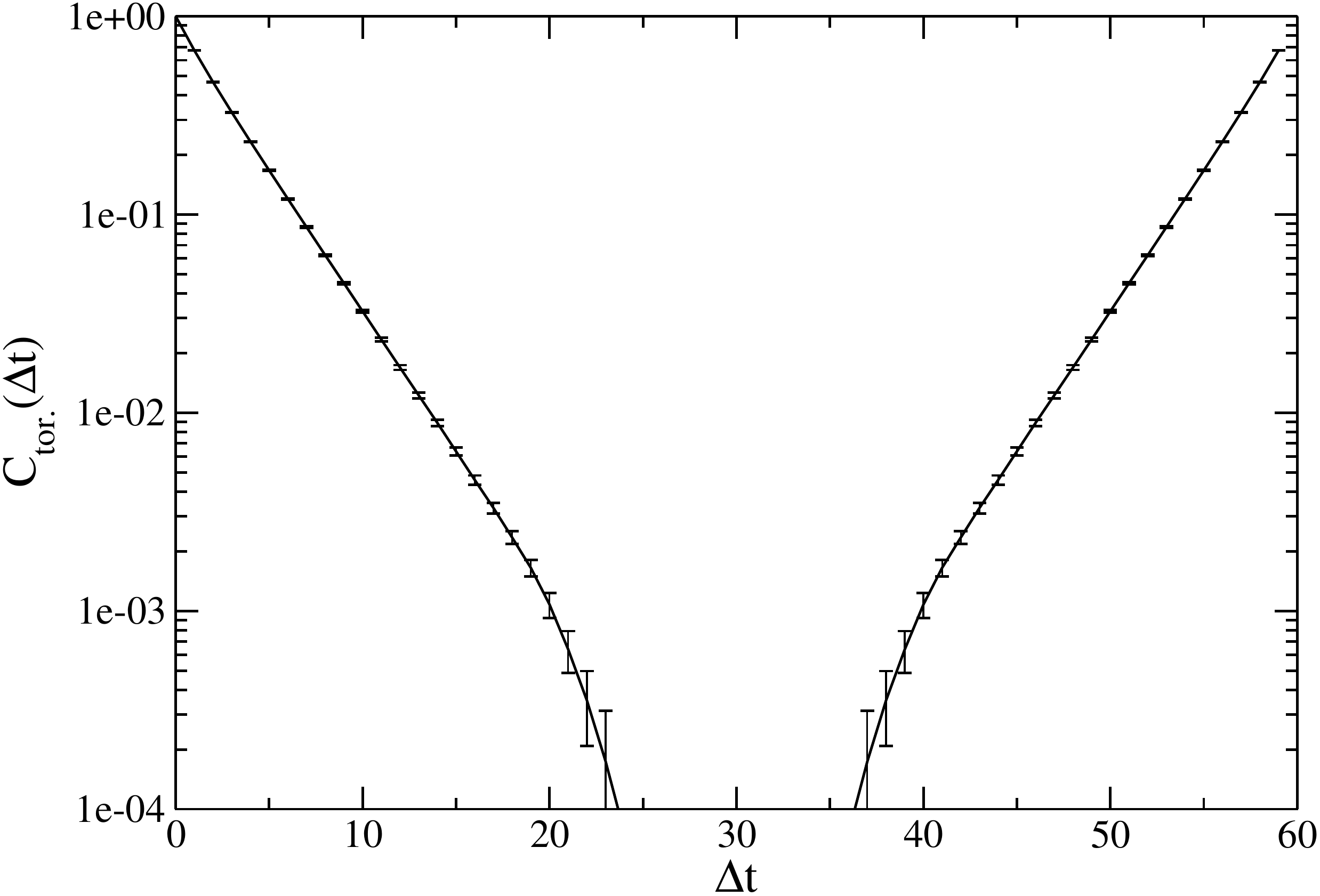}
\caption{Connected two-point correlation function of torelon profiles 
for a simulation at $\lambda=-0.44244$ and $\beta=90$.}
\label{fig:ccorr_torel_36_wmed}
\end{figure}

In order to go ahead with this program one should consider 
a set of different correlation lengths, both gravity and gauge related,
in order to perform a consistent analysis. While we plan to do that
more systematically in future studies, we have considered, as a first
correlation length linked to the gauge field dynamics, 
that related to the 
\emph{torelon profile} $\Pi(t)$, i.e.~the holonomy, 
which is in some sense analogous
to the volume profile:
for a two dimensional torus triangulation 
we define the \emph{torelon}, in analogy with the definition
given on flat lattices~\cite{torelons_ref1,torelons_ref2}, as 
the oriented product of link variables over the closed loop of dual-links that
wraps around along a \emph{slab},
that is a strip of triangles betwen two slices (labeled by $t$ as for slices);
the trace of the torelon is a gauge invariant object.

\begin{figure}
\centering
\includegraphics[width=0.5\textwidth]{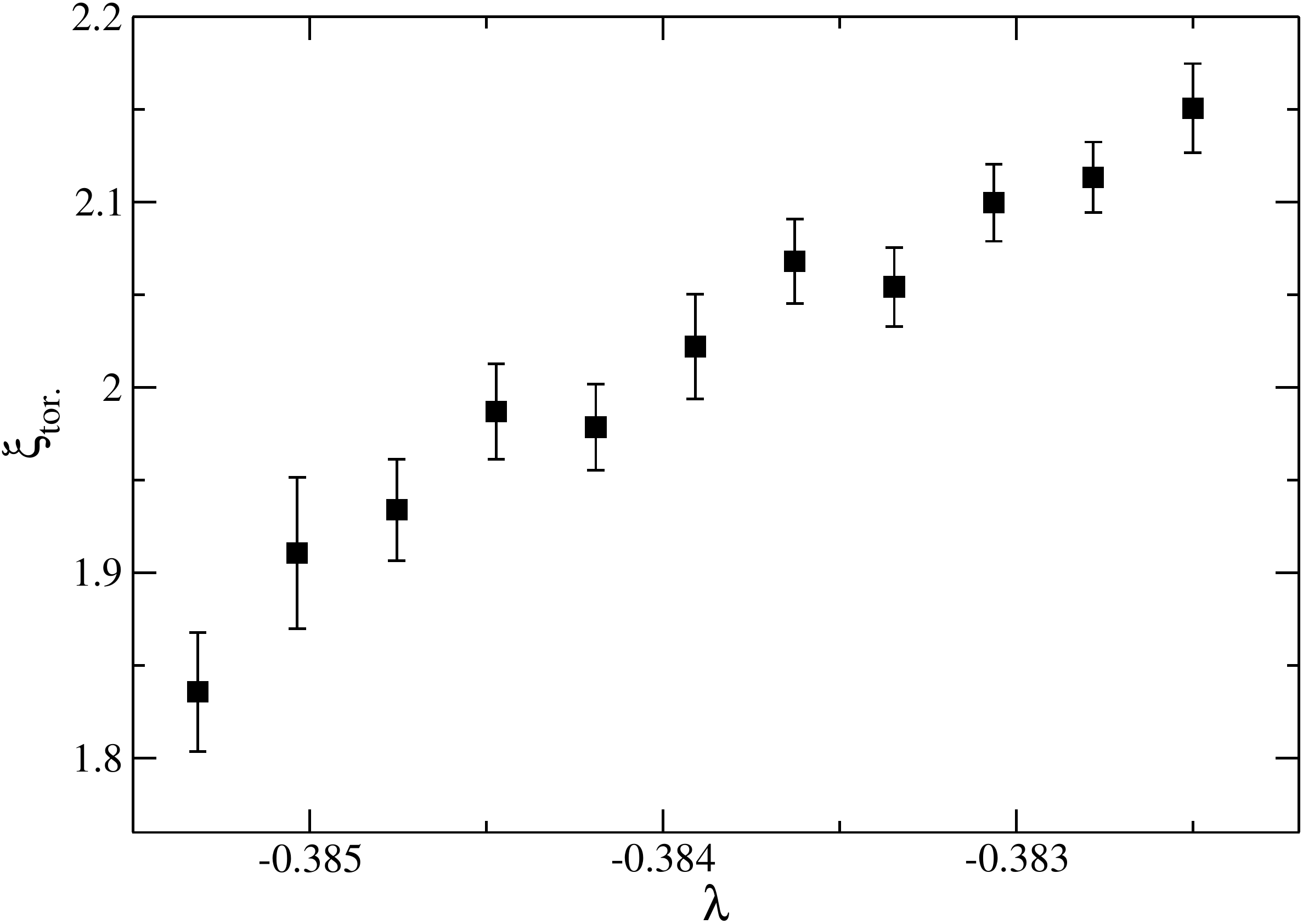}
\caption{Torelon correlation length at $\beta=70$, as a function of the $\lambda$ parameter.}
\label{fig:xi_torel_vs_lambda}
\end{figure}

As for the volume profiles, 
we study the connected two point correlation functions of
torelonic profiles, defined as:
\begin{equation}\label{eq:un-corr2torel}
	C_{\text{tor}}(\Delta t) \equiv
\frac{\langle Tr \Pi(t) Tr \Pi^\dagger(t+\Delta t) \rangle - 
|\langle Tr \Pi(t) \rangle|^2}{\langle |Tr \Pi(t)|^2\rangle - 
|\langle Tr \Pi(t) \rangle|^2} \, .
\end{equation}
The same considerations done for Eq.~\eqref{eq:un-corr2prof} apply here,
and from the exponential decay of the 
correlator we can extract
the \emph{torelon correlation length} $\xi_{\text{tor.}}$.\\
Fig.~\ref{fig:ccorr_torel_36_wmed}
shows a typical torelonic correlation function, while
in Fig.~\ref{fig:xi_torel_vs_lambda} we report
the torelon correlation length measured for 
$\beta = 70$ and variable $\lambda$ (above $\lambda_c(\beta)$). 
Such preliminary results show, once again, that the coupling to gravity has a non-trivial
effect on gauge field dynamics, which would be interesting to better
investigate in the future.

\section{Discussion and Conclusions}\label{sec:conclusions}

Our study has been devoted to the numerical implementation
and investigation of compact gauge fields minimally coupled
to Dynamical Triangulations. We have described a formulation
in which local gauge transformations are associated with
the elementary simplexes of the triangulation, so that the
elementary gauge variables describing the gauge connection
live on the links of the dual graph, which connect adiacent simplexes.

After writing an expression for the discretized plaquette
action, which is reported in Eq.~(\ref{eq:un-YMact_cdt}) and
is valid for a generic dimension and gauge group, we have focused
on two-dimensional Causal Dynamical Triangulations and on
two different choices of the gauge group, either $U(1)$ or $SU(2)$.
For these cases, we have developed a set of Markov chain moves
for the Monte Carlo sampling of the discretized path-integral,
consisting in a standard heat-bath for gauge fields at fixed
triangulation plus the $(2,2)$, $(2,4)$
and $(4,2)$ moves usually adopted in 2D CDT, which
have been properly modified to deal with the presence
of gauge fields while preserving detailed balance
(see the Appendix for a detailed discussion).

As a first numerical analysis, we have considered the critical
behavior of gravity observables, in particular the total volume
$V$ of the triangulation and the correlation length of
the volume profile $V(t)$. In absence of the gauge fields,
both are found to diverge as the cosmological coupling
$\lambda$ approaches a critical value $\lambda_c$ from above,
with $\lambda_c = 0.69319(5)$, in agreement with
the theoretical prediction $\lambda_c = \log 2\simeq 0.693147$. 
The addition of gauge fields with a finite bare gauge coupling
$\beta$ has been found
to modify the above scenario just for a 
shift in $\lambda_c$, which for low $\beta$
has been found to be consistent with a 
strong coupling estimate.
In particular, the critical indices $\mu$ and $\nu$ associated
with the divergence of $\langle V \rangle$ and of the correlation length
are found to be independent of $\beta$ within errors, and 
from a fit to a constant value we have 
obtained $\mu = 0.363(3)$ and $\nu = 0.496(7)$,
i.e.~$\nu$ is compatible with a mean field value.

Our numerical results show also that gauge field dynamics, instead,
is influenced by the coupling to gravity in a less trivial way.
In this exploratory investigation we have considered just torelon
correlators with the associated correlation length and, in the
case of the $U(1)$ gauge group, the dependence on the topological
parameter $\theta$.

An interesting observation regards
a well known problem affecting the numerical simulations
of standard lattice gauge theories, namely the critical slowing
down of topological modes as the continuum limit is approached, also
known as topological freezing. We have found, by direct comparison
with simulations on a flat static lattice, that the problem
is strongly reduced, by orders of magnitude in
terms of the autocorrelation time, in the presence of a dynamical
triangulation. Morover, we have clarified that the improvement is not
linked to the dynamics of the triangulation, but just to
local variations of the geometry, even at fixed triangulation: that
could be due to the presence of \emph{larger holes} around bones
with a negative curvature, through which local topological fluctuations
could happen more easily.
Such observation could have interesting implications also for
algorithmic developments in standard lattice gauge theories,
where the local variation of the geometry could be mimicked
in some way, not necessarily involving a coupling to gravity.

Future developments stemming from this exploratory study are expected
in various directions. First of all, a more systematic
determination of gravity related and gauge related correlation lengths
will permit to fix the curves $\lambda(\beta)$
in the $\lambda-\beta$ plane along which the ratios of correlation
lengths are kept fixed, i.e., the so-called lines of constant physics.
Along these lines, a proper continuum limit could be taken,
with all correlation lengths diverging at the same time, and with
each line corresponding to a different renormalized
quantum field theory.
Another obvious extension is that to a larger number of dimensions,
where the problem of finding a proper continuum limit is
less trivial also in absence of the gauge fields, and where
also the development of the set of Markov chain moves will be
more challenging.

\acknowledgments
 
We thank Claudio Bonati, Paolo Rossi and Francesco Sanfilippo 
for useful discussions. 
Numerical simulations have been performed on the MARCONI machine at CINECA, 
based on the agreement between INFN and CINECA 
(under projects INF19\_npqcd and INF20\_npqcd), 
and at the IT Center of the Pisa University.

\appendix
\section{Detailed balance for the Markov chain Monte Carlo moves}
As mentioned in Section~\ref{sec:algo}, 
the path-integral probability weight 
for a configuration $(\mcTau_i,\Phi_i)\in \mathcal{C}_{G}(\mathcal{T})$
of the composite gravity-gauge system can be written as:
\begin{equation}
    P_i = \frac{1}{Z} e^{- S_{\text{CDT}}[\mcTau_i] - S_{\text{YM}}[\Phi_i;\mcTau_i]},
\end{equation}
where $Z$ is the normalization factor (\emph{partition function}).
 
In order to ensure that the Markov chain of the Monte Carlo simulation converges to the 
correct equilibrium probability distribution $P_i$, we impose 
the detailed balance condition
on the transition probability $W_{i \to f}$ of passing
from an initial configuration $i$ to a final one $f$.
Apart from the pure gauge move, in which 
we adopt a heat-bath algorithm, the other steps 
of the Markov chain are Metropolis--Hastings steps~\cite{metro,hastings}, i.e., they
consist in first selecting
a tentative next configuration $f$ with probability 
$P_{\text{sel.}}(f|i)$ and then accepting it with an acceptance 
ratio $A_{i \to f}$. In the case in which $f$ is not accepted, the final
configuration is taken to be equal to the initial one $i$.
The transition probability can therefore be decomposed as the 
product of the selection probability and the acceptance ratio
$W_{i \to f} = P_{\text{sel.}}(f|i) A_{i \to f}$ and the
detailed balance condition takes the form:
\begin{equation}\label{eq:db-cond}
    P_{\text{sel.}}(f|i) A_{i\rightarrow f} P_i  = P_{\text{sel.}}(i|f) A_{f\rightarrow i} P_f.
\end{equation}
Notice that, in the general case, the selection probability cannot
be controlled and we have to select the acceptance ratio in 
such a way as to reproduce the correct equilibrium distribution.
The optimal values of the $A$s that maximize the acceptance can be found
to be equal to:
\begin{subequations}\label{eq:un-db-general_acceptance_both}
\begin{align}
A_{i\rightarrow f} &= \min \Big( 1, \frac{P_{\text{sel.}}(i|f)}{P_{\text{sel.}}(f|i)} \cdot \frac{P_f}{P_i} \Big),\label{eq:un-db-general_acceptance_if} \\
A_{f\rightarrow i} &= \min \Big( 1, \frac{P_{\text{sel.}}(f|i)}{P_{\text{sel.}}(i|f)} \cdot \frac{P_i}{P_f} \Big).\label{eq:un-db-general_acceptance_fi}
\end{align}
\end{subequations}
In this Appendix, we are going to use 
Eqs.~\eqref{eq:un-db-general_acceptance_both} to derive the 
acceptance probabilities that satisfy the detailed balance condition 
for the adapted Alexander moves $(2,2)$,
$(2,4)$ and $(4,2)$, that have already been schematically described
in Section~\ref{sec:algo}. We stress that, given the complexity
introduced in the moves, we have performed a series of 
preliminary debugging runs in which the moves have been combined
with different weights, in order to check for the correctness of 
each single move by verifying the stability 
of various average observables, like the average volume 
or the average plaquette, against the change in the weights.

\subsection{Detailed balance for the $(2,2)$ move}
As this move does not change the total number of triangles,
the variation of the gravitational part of the action vanishes: 
$\Delta S^{(2,2)}_{\text{CDT}} = 0$.
However, the selection probabilities of the initial and final 
triangulations $P_{\text{sel.},\text{CDT}}[\mcTau]$ and 
$P_{\text{sel.},\text{CDT}}[\mcTau^\prime]$ are in general 
different, giving a non trivial contribution to the acceptance:
we do not review here the computation of such probabilities, 
which is described in standard literature on dynamical triangulations 
(see, e.g., Ref.~\cite{cdt_review12}).

Regarding the gauge field configuration, the gauge fixing 
of the central dual
link variable to the group identity $\mathds{1}$ 
ensures that this move, involving a flip of a time-like link,
does not change the values of the plaquettes but only
their lengths (i.e., the coordination numbers
of the associated vertices), as can be seen 
from Fig.~\ref{fig:move22_draw}, to which we refer 
in the following.
Since the YM gauge action~\eqref{eq:un-YMact_cdt} 
depends explicitly on the plaquette lengths, its variation
is non trivial:
\begin{align}\label{eq:un-move22-actionGvar}
    &\Delta S_{\text{YM}}^{(2,2)} =\\& \beta\Big[ \frac{\widetilde{\Pi}_1}{n_1 (n_1+1)} + \frac{\widetilde{\Pi}_2}{n_2 (n_2+1)} - \frac{\widetilde{\Pi}_3}{n_3 (n_3-1)} - \frac{\widetilde{\Pi}_4}{n_4 (n_4-1)}\Big],\nonumber
\end{align}
where we report the definition \mbox{$\widetilde{\Pi}_b\equiv \lbrack \frac{1}{N} Re Tr \Pi_b - 1 \rbrack$}.
We also stress that the selection probabilities of the initial
and final gauge configuration $\Phi$ and $\Phi^\prime$ 
are trivially equal, 
$P_{\text{sel.},\text{YM}}[\Phi] = 
P_{\text{sel.},\text{YM}}[\Phi^\prime]$,
because, given the gauge fixing, the selection
procedure is deterministic in both directions.

Finally, the acceptance ratio can be found from 
Eqs.~\eqref{eq:un-db-general_acceptance_both}, receiving 
contributions both from the non-trivial ratio of the selection 
probabilities of the triangulations and the variation of the 
gauge action:
\begin{equation}\label{eq:un-move22-acceptprob}
	A^{(2,2)} = \min \Big( 1, e^{-\Delta S_{\text{YM}}^{(2,2)}} \cdot \frac{P_{\text{sel.},\text{CDT}}[\mcTau]}{P_{\text{sel.},\text{CDT}}[\mcTau^\prime]} \Big).
\end{equation}
\subsection{Detailed balance for moves $(2,4)$ and $(4,2)$}\label{subsec:un-db_move2442}
As we will see, this pair of moves is the most involved.
To fix the ideas, let us consider the direct move $(2,4)$, 
which creates two new triangles, increasing the gravitational
contribution to the total action: 
$\Delta S^{(2,4)}_{CDT} = - \Delta S^{(4,2)}_{CDT} = 2 \lambda$.
The effect on the gauge configuration is the introduction
of a new plaquette of length four, which can always be made
to have all link variables but one gauge fixed to the identity $\mathds{1}$.
The remaining link variable have to be extracted from the gauge group $G$
and cannot be set to the identity, since in the inverse move 
$(4,2)$ the plaquette which is eliminated has in general a non-trivial
value, i.e., is not fixed to the identity.
In the following we make reference to Fig.~\ref{fig:move2442_draw} for
the notation.

Let $U^\prime$ denote the newly extracted link variable.
The variation of the gauge action receives 
contributions from both the change in the 
lengths of the plaquettes $\Pi_0$, $\Pi_3$ and $\Pi_4$ and from
the introduction of $U^\prime$:
\begin{align}
    \Delta S^{(2,4)}_{\text{YM}} = \beta &\Big\{ \frac{\widetilde{\Pi}_3}{n_3 (n_3+1)} + \frac{\widetilde{\Pi}_4}{n_4 (n_4+1)} + \frac{\widetilde{\Pi}_2}{n_2} + \frac{1}{4} - \frac{1}{n_2} \nonumber\\
                                          &-\frac{1}{N} Re Tr \Big[U^\prime \Big( \frac{\mathds{1}}{4} + \frac{X_2^\dagger}{n_2}\Big)\Big] \Big\},\label{eq:un-move24-deltaSgauge}
\end{align}
where $X_2$ is the staple of $U^\prime$ associated with the 
plaquette $\Pi_2^\prime$.

There is a way to extract $U^\prime$ which 
is more convenient from the point of view of the acceptance probability:
this represents a non-trivial gain, 
given the complexity of even trying the $(2,4)$
move or its inverse.
We start noticing that, 
in Eq.~\eqref{eq:un-move24-deltaSgauge}, we can separate 
the term dependent on the new link variable $U^\prime$ from the one independent of it, as follows:
\begin{equation}
\Delta S^{(2,4)}_{YM}(U^\prime, \{\Pi_b\}) = \widetilde{\Delta S}^{(2,4)}_{YM}(U^\prime, \{\Pi_b\}) + \widehat{\Delta S}^{(2,4)}_{YM}(\{\Pi_b\}), 
\end{equation}
where we have defined:
\begin{subequations}
\begin{align}
    \widetilde{\Delta S}^{(2,4)}_{YM}(U^\prime, \{\Pi_b\}) = &-\frac{\beta}{N} Re Tr \Big[ U^\prime F^\dagger \Big],\\
    \widehat{\Delta S}^{(2,4)}_{YM}(\{\Pi_b\}) = &+\beta \Big[ \frac{\widetilde{\Pi}_3}{n_3 (n_3+1)} + \frac{\widetilde{\Pi}_4}{n_4 (n_4+1)} \nonumber \\
                                                 & + \frac{\widetilde{\Pi}_2}{n_2} + \frac{1}{4} - \frac{1}{n_2}\Big],
\end{align}
\end{subequations}
and the force acting on the new link variable $U^\prime$ 
is equal to \mbox{$F\equiv \mathds{1}/4 + X_2/n_2$}.

The idea is to extract $U^\prime$ according to its distribution
in the heat bath fixed by the force $F$, 
i.e., with probability distribution:
\beq
\frac{d U^\prime}{Z_G(F)} \exp \left( \frac{\beta}{N} \textrm{Re Tr} 
\left[ U^{\prime} F^{\dagger} \right] \right) \, ,
\label{eq:2442sel}
\eeq
where we have written the normalization factor explicitly
\beq
Z_G(F) \equiv \bigintsss\limits_{G} d U^\prime \exp \left( \frac{\beta}{N} \textrm{Re Tr} 
\left[ U^{\prime} F^{\dagger} \right] \right) \, .
\eeq
In this way, since 
the initial configuration has a unique gauge equivalence class,
the gauge part of the ratio of selection probabilities turns
out to be exactly equal to Eq.~(\ref{eq:2442sel}), and the 
final Metropolis--Hastings acceptance probability 
gets partially simplified, in particular we obtain  
\begin{subequations}
\begin{align}\label{eq:un-move24-acceptprob}
    A^{(2,4)} &= \min \Big\{ 1, \frac{P_{\text{sel.},\text{CDT}}(\mcTau)}{P_{\text{sel.},\text{CDT}}(\mcTau^\prime)} e^{-2 \lambda - \widehat{\Delta S}^{(2,4)}} Z_G(F) \Big\},\\
    A^{(4,2)} &= \min \Big\{ 1, \frac{P_{\text{sel.},\text{CDT}}(\mcTau^\prime)}{P_{\text{sel.},\text{CDT}}(\mcTau)} e^{+2 \lambda + \widehat{\Delta S}^{(2,4)}}/Z_G(F) \Big\}.
\end{align}
\end{subequations}
Also in this case we do not discuss in detail the ratio 
of selection probabilities related to the triangulation itself,
i.e., ${P_{\text{sel.},\text{CDT}}(\mcTau)}/{P_{\text{sel.},\text{CDT}}(\mcTau^\prime)}$, which is described in previous literature~\cite{cdt_review12}.

Notice that
the acceptance probability depends explicitly on the normalization factor
$Z_G(F)$ computed as a Haar integral. 
As the force $F$ changes dynamically,
it is necessary to compute the normalization function during the 
simulation. 
For certain groups, like $U(N)$ and $SU(N)$, an analytic expression
for this integral is known~\cite{Zintegral1,Zintegral2}. In particular, for the groups
$U(1)$ and $SU(2)$ that we have studied in the numerical 
investigation, the normalization functions are given by:
\begin{align}
Z_{U(1)}(F) &= I_0 \Big(2 det(J)\Big); \\
Z_{SU(2)}(F) &= \frac{1}{\sqrt{K(J)}} I_1 \Big( 2 \sqrt{K(J)}\Big), 
\end{align}
where $J\equiv \frac{1}{2 g^2 N} F$,
\mbox{$K(J)\equiv Tr\big( J^\dagger J \big) + det(J) + det(J^\dagger)$}, 
and we have used the modified Bessel functions of the first kind,
$I_k$.


\begin{thebibliography}{99}

 \bibitem{ass_weinberg}
    S.~Weinberg,
    ``General Relativity, an Einstein Centenary Survey'', 
    ch.16 Cambridge Univ. Press (1979).

 \bibitem{regge} 
    T.~Regge,
    Nuovo Cim.\ {\bf 19} (1961), 558.

\bibitem{cdt_pioneer}
J.~Ambjorn, J.~Jurkiewicz and R.~Loll,
Phys. Rev. Lett. \textbf{85}, 924-927 (2000)
[arXiv:hep-th/0002050 [hep-th]].

 \bibitem{cdt_review12} 
	J.~Ambjorn, A.~Goerlich, J.~Jurkiewicz and R.~Loll,
	Phys.\ Rept.\  {\bf 519} (2012), 127
	[arXiv:1203.3591 [hep-th]].

 \bibitem{cdt_review19}
    R.~Loll,
    Class. Quant. Grav. \textbf{37}, no.1, 013002 (2020)
    [arXiv:1905.08669 [hep-th]].


\bibitem{DynTriLor}
J.~Ambjorn, J.~Jurkiewicz and R.~Loll,
Nucl. Phys. B \textbf{610}, 347-382 (2001)
[arXiv:hep-th/0105267 [hep-th]].

\bibitem{cdt_secondord}
J.~Ambjorn, S.~Jordan, J.~Jurkiewicz and R.~Loll,
Phys. Rev. Lett. \textbf{107}, 211303 (2011)
[arXiv:1108.3932 [hep-th]].

\bibitem{cdt_secondfirst}
J.~Ambjorn, S.~Jordan, J.~Jurkiewicz and R.~Loll,
Phys. Rev. D \textbf{85}, 124044 (2012)
[arXiv:1205.1229 [hep-th]].

\bibitem{new_phase_chars}
J.~Ambj\o{}rn, J.~Gizbert-Studnicki, A.~G\"orlich, J.~Jurkiewicz, N.~Klitgaard and R.~Loll,
Eur. Phys. J. C \textbf{77}, no.3, 152 (2017)
[arXiv:1610.05245 [hep-th]].

\bibitem{cdt_newhightrans}
J.~Ambjorn, D.~Coumbe, J.~Gizbert-Studnicki, A.~Gorlich and J.~Jurkiewicz,
Phys. Rev. D \textbf{95}, no.12, 124029 (2017)
[arXiv:1704.04373 [hep-lat]].


\bibitem{cdt_toroidal}
J.~Ambj\o{}rn, J.~Gizbert-Studnicki, A.~G\"orlich, K.~Grosvenor and J.~Jurkiewicz,
Nucl. Phys. B \textbf{922}, 226-246 (2017)
[arXiv:1705.07653 [hep-th]].

\bibitem{cdt_quantum_Ricci_curv}
N.~Klitgaard and R.~Loll,
Phys. Rev. D \textbf{97}, no.4, 046008 (2018)
[arXiv:1712.08847 [hep-th]].


\bibitem{cdt_phasestruct_toroidal}
J.~Ambj\o{}rn, J.~Gizbert-Studnicki, A.~G\"orlich, J.~Jurkiewicz and D.~N\'emeth,
JHEP \textbf{06}, 111 (2018)
[arXiv:1802.10434 [hep-th]].

\bibitem{LBseminal}
G.~Clemente and M.~D'Elia,
Phys. Rev. D \textbf{97}, no.12, 124022 (2018)
[arXiv:1804.02294 [hep-th]].


\bibitem{LBrunning}
G.~Clemente, M.~D'Elia and A.~Ferraro,
Phys. Rev. D \textbf{99}, no.11, 114506 (2019)
[arXiv:1903.00430 [hep-th]].

\bibitem{cdt_higherord_toroidal}
J.~Ambj\o{}rn, G.~Czelusta, J.~Gizbert-Studnicki, A.~G\"orlich, J.~Jurkiewicz and D.~N\'emeth,
JHEP \textbf{05}, 030 (2020)
[arXiv:2002.01051 [hep-th]].

\bibitem{cdt_quantum_Ricci_curv_round}
N.~Klitgaard and R.~Loll,
[arXiv:2006.06263 [hep-th]].

\bibitem{LBFEMseminal}
F.~Caceffo and G.~Clemente,
[arXiv:2010.07179 [hep-lat]].

 \bibitem{causconds}
    E.~Minguzzi and M.~Sanchez,
    [arXiv:gr-qc/0609119 [gr-qc]].

 \bibitem{cdtgauge_anal} 
    J.~Ambjorn and A.~Ipsen,
    Phys.Rev.D {\bf 88} (2013), 6
    [arXiv:1305.3148 [hep-th]].

 \bibitem{cdtgauge_noncompact}
    J.~Ambjorn, K.~N.~Anagnostopoulos and J.~Jurkiewicz,
    JHEP \textbf{08} (1999), 016
    [arXiv:hep-lat/9907027 [hep-lat]].


\bibitem{alexander}
    J.W. Alexander,
    Ann. Mat. 31 (1931) 292.

 \bibitem{cdt_2dmoves}
    J.~Ambj\o{}rn, J.~Jurkiewicz and R.~Loll,
    NATO Sci. Ser. C \textbf{556}, 381-450 (2000)
    [arXiv:hep-th/0001124 [hep-th]].

 \bibitem{metro}
    N.~Metropolis, A.~W.~Rosenbluth, M.~N.~Rosenbluth, A.~H.~Teller and E.~Teller,
    J. Chem. Phys. \textbf{21}, 1087-1092 (1953)

 \bibitem{hastings}
    W.~K.~Hastings,
    Biometrika \textbf{57}, 97-109 (1970)


\bibitem{hb_creutz}
M.~Creutz,
Phys. Rev. D \textbf{21}, 2308-2315 (1980)

\bibitem{hb_kennedy-pendleton}
A.~D.~Kennedy and B.~J.~Pendleton,
Phys. Lett. B \textbf{156}, 393-399 (1985)

\bibitem{two_dim_scaling}
    J.~Ambjorn and R.~Loll,
    Nucl. Phys. B \textbf{536}, 407-434 (1998)
    [arXiv:hep-th/9805108 [hep-th]].

\bibitem{two_dim_scaling2}
J.~Ambjorn, R.~Loll, J.~L.~Nielsen and J.~Rolf,
Chaos Solitons Fractals \textbf{10}, 177-195 (1999)
[arXiv:hep-th/9806241 [hep-th]].


\bibitem{Cao:2013na}
C.~Cao, M.~van Caspel and A.~R.~Zhitnitsky,
Phys. Rev. D \textbf{87}, no.10, 105012 (2013)
[arXiv:1301.1706 [hep-th]].

\bibitem{bonati_flatsuscu1}
C.~Bonati and P.~Rossi,
Phys. Rev. D \textbf{99}, no.5, 054503 (2019)
[arXiv:1901.09830 [hep-lat]].


\bibitem{Alles:1996vn}
B.~Alles, G.~Boyd, M.~D'Elia, A.~Di Giacomo and E.~Vicari,
Phys. Lett. B \textbf{389}, 107-111 (1996)
[arXiv:hep-lat/9607049 [hep-lat]].

\bibitem{top_freeze}
L.~Del Debbio, G.~M.~Manca and E.~Vicari,
Phys. Lett. B \textbf{594}, 315-323 (2004)
[arXiv:hep-lat/0403001 [hep-lat]].

\bibitem{Luscher:2011kk}
M.~Luscher and S.~Schaefer,
JHEP \textbf{07}, 036 (2011)
[arXiv:1105.4749 [hep-lat]].


\bibitem{Bonati:2017woi}
C.~Bonati and M.~D'Elia,
Phys. Rev. E \textbf{98}, no.1, 013308 (2018)
doi:10.1103/PhysRevE.98.013308
[arXiv:1709.10034 [hep-lat]].


\bibitem{torelons_ref1}
C.~Michael,
Phys. Lett. B \textbf{232}, 247-250 (1989)

\bibitem{torelons_ref2}
    E.~Marinari, M.~L.~Paciello and B.~Taglienti,
    Int. J. Mod. Phys. A \textbf{10}, 4265-4310 (1995)
    [arXiv:hep-lat/9503027 [hep-lat]].

\bibitem{Zintegral1}
R.~Brower, P.~Rossi and C.~I.~Tan,
Nucl. Phys. B \textbf{190}, 699-718 (1981)

\bibitem{Zintegral2}
R.~C.~Brower, P.~Rossi and C.~I.~Tan,
Phys. Rev. D \textbf{23}, 942 (1981)
 
 \end{thebibliography}
 \end{document}